\documentclass[aps,pra,eps,twocolumn]{revtex4}
\usepackage{graphicx}
\usepackage{subfigure}
\usepackage{bm}

\usepackage[intlimits]{amsmath} 

\usepackage{verbatim} 

\usepackage{amssymb}
\newcommand{\gamm}{\frac{\gamma}{2}}
\newcommand{\bra}[1]{\langle #1  | }
\newcommand{\ket}[1]{ | #1 \rangle  }

\newcommand{\V}[1]{\mbox{\boldmath $#1$}}

\newcommand{\D}[1]{\left ( \V{D}\cdot \V{\epsilon}_{#1} \right )}

\newcommand{\partiell}[2]{\frac{\partial^{#2}}{\partial {#1}^{#2}}}

\newcommand{\sg}{\sigma}
\newcommand{\Av}[1]{  \langle \hspace{0.1cm} #1 \hspace{0.1cm} \rangle  }

\begin{document}

\title{Photon-Mediated Interaction between Two Distant Atoms}

\author{Stefan Rist,$^{1}$ J\"urgen Eschner,$^2$ Markus
Hennrich,$^2$ and Giovanna Morigi$^1$}

\affiliation{
$^1$ Departament de F{\'i}sica, Universitat Aut\`onoma de Barcelona, 08193 Bellaterra, Spain\\
$^2$ ICFO -- Institut de Ci\`encies Fot\`oniques, 08860 Castelldefels (Barcelona), Spain}

\begin{abstract}
We study the photonic interactions between two distant atoms which are coupled by an
optical element (a lens or an optical fiber) focussing part of their emitted radiation
onto each other. Two regimes are distinguished depending on the ratio between the
radiative lifetime of the atomic excited state and the propagation time of a photon
between the two atoms. In the two regimes, well below saturation the dynamics exhibit
either typical features of a bad resonator, where the atoms act as the mirrors, or
typical characteristics of dipole-dipole interaction. We study the coherence properties
of the emitted light and show that it carries signatures of the multiple scattering
processes between the atoms. The model predictions are compared with the experimental
results in J. Eschner {\it et al.}, Nature {\bf 413}, 495 (2001). \end{abstract}

\date{\today}
\maketitle

\section{Introduction}

Control of photon-atom interaction lies at the heart of quantum technologies based on
atomic and photonic systems~\cite{ZollerRoadmap}. Recent experiments demonstrated the
quantum correlations between atoms and emitted
photons~\cite{Monroe04,Weinfurter06,Grangier06}. Atom-photon entanglement was then
applied for entangling distant atoms by photon measurement~\cite{Maunz}. Further
experiments demonstrated the possibility to spatially confine atoms with nanometric
precision inside resonators~\cite{Guthoerlein2001,Mundt,Rauschenbeutel}, and hence to
control their coupling with the electromagnetic field modes of cavities. Such precision
has permitted realizing quantum light sources with high degree of
control~\cite{Kimble04,Walther04,Kuhn2,Rempe07,Kimble08}, and hence to pose the basis for
the realization of quantum networks based on atom-photon interfaces~\cite{ZollerRoadmap}.

Parallel to these experimental efforts, studies are also focussing on achieving strong
coupling between atoms and photons by means of optical elements, such as lenses of large
numerical aperture~\cite{Eschner2001, Wilson, Bushev2004, Sondermann07, Leuchs07,
Kurtsiefer} or optical fibers~\cite{Spreeuw,Balykin,Arno}. In particular,
in~\cite{Eschner2001} two distant atoms in front of a mirror were coupled by means of a
lens, focussing the radiation emitted by one atom onto the other. In this setup, the
first-order coherence was experimentally studied, showing an interference pattern when
the optical path length between the atoms was varied. In earlier experiments with two
trapped ions, far-field interference of their scattered light~\cite{Itano}, and their
near-field interaction~\cite{DeVoe} were studied.

\begin{figure}[htp]
     \centering
\includegraphics[width=.3\textwidth]{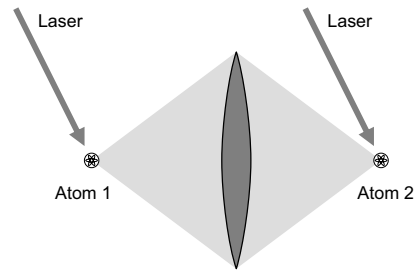} \caption{The
dipolar transitions of two atoms, which are several optical
wavelength apart, are coupled by a lens focussing part of their
emitted radiation onto each other. In~\cite{Eschner2001} a similar
situation was realized, coupling two atoms via a mirror and a
lens. Analogous dynamics can be observed when the atoms are
trapped close to an optical fiber, see for instance~\cite{Arno}.}
     \label{Fig:1}
\end{figure}

In this article we present an extensive theoretical study of the radiative properties of
two distant atoms when they are coupled via an optical element, which could be an optical
fiber or a lens, as sketched in Fig.~\ref{Fig:1}. In this situation radiation is multiply
scattered between the atoms, until it is finally dissipated into the external modes of
the electromagnetic field. Our model is based on the theory developed
in~\cite{Alber,Dorner} for the case of a single atom interacting with itself via a
mirror, and extends it to the situation of two coupled atoms. The theoretical predictions
of our model reproduce the experimental results of~\cite{Eschner2001} and allow us to
identify possible measurements that highlight the multiple-scattering features. Moreover,
the scattered photons are correlated with the scattering atoms, thereby establishing
correlations and, in certain cases, entanglement between their internal excitations.

This article is organized as follows. In Sec.~\ref{Sec:Summary} we
make some preliminary considerations on the system. In
Sec.~\ref{Sec:Model} we introduce the model in detail and solve
the basic equations describing the coupled dynamics of the
internal atomic states and few photons of the electromagnetic
field. In Sec.~\ref{Sec:NoLaser} we investigate in detail the
radiative properties of the system, and in Sec.~\ref{Sec:Laser} we
provide the details of the first- and second-order coherence of
the light scattered by the atoms when they are weakly driven by a
laser. In Sec.~\ref{Sec:Conclusion} we provide some outlooks to
the present work, and in the appendices we report details of the
calculations.

\section{Preliminary considerations}
\label{Sec:Summary}

The scattering cross section of an atomic dipole transition in free space is on the order
of the square of its wavelength $\lambda$~\cite{Cohen-Tannoudji}. Consequently, the
free-space photonic interaction between two atomic dipoles at distance $d$ is determined
by the ratio $\lambda/d$~\cite{Dipole-Dipole}: when $d\ll\lambda$,
strong modifications of the atomic emission spectrum of one atom due to the presence of
another one are observable~\cite{Dipole-Dipole, DeVoe, Savels07};
when $d\gg\lambda$, these effects are negligible,
and the atoms scatter photons independently. This behavior is dramatically modified if an
optical system, like a lens with large numerical aperture or an optical fiber, focusses a
significant fraction of the radiation
emitted by one atom onto the other. This latter situation is sketched in Fig.~\ref{Fig:1}
for the case of a lens that images the atoms onto each other.

When the photonic interaction between the atoms is mediated by an optical element, its
strength is characterized by the fraction $\kappa$ of modes of the electromagnetic field
which the optical system transforms into each other. Thus $\kappa$ replaces the scaling
with $\lambda/d$ of the free-space case, and coupling over much larger distances than
$\lambda$ may be achieved.

The atom-atom distance $d$, or more precisely the propagation time
for a photon from one atom to the other via the optical element,
\begin{equation}\label{tau} \tau=\frac{d}{c} \end{equation}
remains an important physical parameter of the photonic
interaction, since it has to be compared with the radiative
lifetime of the atomic dipolar transition $1/\gamma$, which
determines the time scale on which the photonic excitation is
dissipated into free space, as well as the length of the emitted
photonic wave packet. When $\gamma\tau\gg 1$, the process of
photon scattering by each atom is well localized in time and
space: a photonic excitation is exchanged between the atoms at
integer multiples of the delay time $\tau$, until its amplitude is
damped to zero by emission into the external modes of the
electromagnetic field. When $\gamma\tau\ll 1$, in contrast,
multiple scattering events add up coherently during the excitation
time of each atom, causing the spontaneous emission rate to be
enhanced or suppressed, depending on the interatomic distance
(modulo the wavelength). This regime is equivalent to
dipole-dipole interaction with a delay time $\tau$.

In all cases, the system of two atoms confining radiation by multiple scattering shows
some analogies with an optical resonator with low-reflectivity mirrors. This analogy is
appropriate when the atomic transition is not saturated. Indeed, in this regime the
radiative properties are very similar to those of a single atom interacting with itself
via a mirror, studied in~\cite{Eschner2001, Dorner, Wilson}. The peculiarity of the
two-atom system becomes more evident when saturation effects are relevant. Some important
properties, such as the creation of correlations and entanglement between the atoms via
the multiply scattered photons, are identified when studying intensity-intensity
correlation of the light scattered by the two-atom system, as discussed in
Sec.~\ref{Sec:Laser}.

\section{The Model} \label{Sec:Model}

In this section we develop the theoretical model for describing the dynamics of two atoms
in presence of an optical element which focusses the radiation emitted by each atom into
the other, as sketched in Fig.~\ref{Fig:1}. In particular, we use the
theoretical formalism in~\cite{Dorner} for one atom in front of a mirror, and generalize
it to the case of two coupled atoms.

The system consists of two identical atoms of mass $M$, which are trapped at the
positions ${\bf r_1}$ and ${\bf r_2}$, and whose relevant electronic degrees of freedom
are the ground state $|g\rangle$ and the excited state $|e\rangle$ forming a dipole
transition with dipole moment $\V{D}$, frequency $\omega_0$ and wavelength $\lambda=2\pi
c/\omega_0$. The interatomic distance $d=|{\bf r_2}-{\bf r_1}|$ is such that
$d\gg\lambda$, thus free-space dipole-dipole interaction between the atoms is negligible.
We assume, however, that a lens (or an equivalent optical system) is placed between the
atoms, which collects a fraction of the radiation from each atom and focusses it onto the
other one. We use the plane wave decomposition for these modes and label them with
$\rho$, in order to distinguish them from the external modes which do not couple the
atoms; the latter are labeled with $\mu$, see also Fig.~\ref{fig:FullSetup}. The Hamiltonian of the system describes the
interaction between the dipoles and the modes of the electromagnetic field, and can be
decomposed into the sum
\begin{equation}\label{eqn:Hamiltonian}
H=H_0+V_{\rm emf}\,,\end{equation}
where $H_0$ gives the self-energy and $V_{\rm emf}$ the interaction between the dipoles
and the modes of the electromagnetic field. In detail,
\begin{equation}
\label{Hamilton0} H_0= \sum_{j=1,2} \hbar \omega_0
\sigma^+_j\sigma^-_j +\sum_{\ell=\rho,\mu} \hbar \omega_{\ell}
a_{\ell}^{\dagger} a_{\ell}\,,
\end{equation}
where the first term describes the energy of the atoms, with $\sigma_j=|g\rangle_j
\langle e|$ and $\sigma_j^{\dagger}$ its adjoint, and subscript $j=1,2$ labeling the
atom. The second term is the free Hamiltonian of the transverse photon field where the
summation runs over all field modes. We label by $\ell$ the mode with wave vector ${\bf
k_{\ell}}$ and polarization $\V{\epsilon}_{\ell} \perp {\bf k_{\ell}}$, while
$a_{\ell}^{\dagger}$ and $a_{\ell}$ are the creation and annihilation operators for a
photon in that mode, obeying the commutation relation $\left[
a_{\ell},a_{\ell'}^{\dagger} \right] = \delta_{\ell,\ell'}$. In particular, the modes
with label $\ell=\rho$ are the ones which couple the atoms via the lens.

The interaction of the atoms with the electromagnetic field, $V_{\rm emf}$, is given in
the electric dipole and rotating wave approximation, and takes the form
\begin{eqnarray} \label{eqn:2AV}
V_{\rm emf} &=& -{\rm i}\hbar \sum_{j=1,2} \sigma_j^+ \sum_{\ell=\rho,\mu} g_{\ell}
a_{\ell} {\rm e}^{{\rm i}{\bf k_{\ell}}\cdot {\bf r_j}}+{\rm H.c.}\,,
\end{eqnarray}
where $$g_{\ell} = \D{\ell} \sqrt{\omega_{\ell}/(2 \varepsilon_0 \hbar {\mathcal V})}$$
with the vacuum electric permittivity $\varepsilon_0$ and the quantization
volume~$\mathcal V$.

In presence of a laser driving the atoms the Hamiltonian will be given by
\begin{equation}\label{H:tot} H'=H+V_L(t)\,,\end{equation} where the term $V_L$ describes
the atom-laser coupling and reads
\begin{equation}\label{Vlaser} V_L =\hbar \Omega
\sum_{j=1,2}\sigma^+_j {\rm e}^{{\rm i}({\bf k_L}\cdot {\bf r_j}-\omega_Lt)} + {\rm H.c.}
\end{equation} Here, the laser is a classical field at frequency
$\omega_L$~\cite{Cohen-Tannoudji}, $\Omega$ is the coupling strength, and ${\bf k_L}$ is
the wave vector of the incident laser beam.

The dynamics of the system is studied by solving the Schr\"odinger equation treating the
interaction of the atoms with the electromagnetic field as a perturbation. For this
purpose, the wave function $\ket{\psi(t)}$ of the atoms and the field at time $t$, in the
interaction picture with respect to $H_0$, is described by
\begin{eqnarray}\label{eqn:wavefct}
&&\ket{\psi (t)}=b_e^{(1)}(t)
\ket{e_1,g_2,0}+ b_e^{(2)}(t) \ket{g_1,e_2,0}\\
&&+\sum_{\rho} b_g^{(\rho)}(t) \ket{g_1,g_2,1_{\rho},0_{\mu}}
+\sum_{\mu} b_g^{(\mu)}(t) \ket{g_1,g_2,0_{\rho},1_{\mu}}\,,
\nonumber\end{eqnarray}
where the state $\ket{0}$ corresponds to the vacuum state of the electromagnetic field,
and the state $\ket{n_\mu}$ ($\ket{n_{\rho}}$) to $n$ photons in mode $\mu$ ($\rho$). In
Eq.~(\ref{eqn:wavefct}) we have assumed that at most one excitation is present in the
system. In particular, the coefficients $b_e^{(j)}(t)$ are the probability amplitudes at
time $t$ for atom $j$ being in the excited state, while the coefficient $b_g^{(\ell)}(t)$
gives the probability amplitude to find a photon in the field mode $\ell$ at time $t$,
with both atoms in the ground state. For later convenience, we also introduce the
probability amplitudes $b_g^{(j,\ell)}(t)$, with
$$b_g^{(\ell)}(t)=b_g^{(1,\ell)}(t)+b_g^{(2,\ell)}(t)\,,$$ and which distinguish which atom
has emitted the photon into mode $\ell$.

We will solve the Schr\"odinger equation using this ansatz first in absence and then in
presence of a laser driving the atoms. In particular, we will study the dynamics as a
function of two important physical quantities which characterize the system. The first is
the time delay $\tau$ for light to propagate from one atom to the other, defined in
Eq.~(\ref{tau}). As noted before, we consider the case $c\tau\gg \lambda$. The second
important quantity is the strength of the photonic coupling between the atoms mediated by
the lens, which is defined through the fraction of $4\pi$ solid angle within which the
radiation from one atom is focussed onto the other. This corresponds to the fraction of
modes labeled with $\rho$, which propagate from one atom to the other via the lens. We
denote the coupling by the dimensionless parameter $\kappa$,
\begin{eqnarray} \label{kappa}
\kappa &=& \sum_{\bf n_{\rho}}
\left(1-|{\bf D}\cdot {\bf n_{\rho}}|^2/|{\bf D}|^2\right)\nonumber\\
&=& \frac{3}{8\pi}\int_{\delta\Omega} d\Omega_0 \left (1-|{\bf D}\cdot {\bf n}|^2/|{\bf
D}|^2 \right )\,,
\end{eqnarray}
where ${\bf n_{\rho}}={\bf k_{\rho}}/k$ and $\delta \Omega$ is the solid angle collected
by the lens. The value of $\kappa$ lies in the interval $0< \kappa <1$, whereby
$\kappa\to 0$ corresponds to the limit without the lens and $\kappa\to 1$ would describe
an ideal optical system that maps all radiation from one atom onto the other.

\subsection{Perturbative solution of the Schr\"odinger equation in
absence of the laser.} \label{Sec:Eqs:NoLaser}

When the atom-laser coupling is set to zero, then in the reference frame of the atoms the
coefficients $b_{e}^{(j)},b_g^{(j,\rho)},b_g^{(j,\mu)}$ obey the differential equations
\begin{subequations} \label{eqn:koeff}
\begin{eqnarray} \dot{b}_{e}^{(j)}(t) &=&
-\sum_{\rho} g_{\rho} {\rm e}^{{\rm i}{\bf k_{\rho}}\cdot {\bf r_j}}
{\rm e}^{{\rm i} (\omega_0 -\omega_{\rho})t} b_g^{(\rho)}(t) \nonumber \\
 && -\sum_{\mu} g_{\mu} {\rm
e}^{{\rm i}{\bf k_{\mu}}\cdot {\bf r_j}}
{\rm e}^{{\rm i} (\omega_0 -\omega_{\mu})t} b_g^{(j,\mu)}(t)\,,\label{1} \\
\dot{b}_g^{(j,\rho)}(t) &=& \, g_{\rho}{\rm e}^{-{\rm i}{\bf
k_{\rho}}\cdot {\bf r_j}}
{\rm e}^{-{\rm i} (\omega_0 -\omega_{\rho})t} b_e^{(j)}(t)\,,\label{eqn:koefffield:rho}\\
\dot{b}_g^{(j,\mu)}(t) &=& \,\, g_{\mu} {\rm e}^{-{\rm i}{\bf
k_{\mu}}\cdot {\bf r_j}} {\rm e}^{-{\rm i} (\omega_0
-\omega_{\mu})t} b_e^{(j)}(t)\,, \label{eqn:koefffield}
\end{eqnarray} \end{subequations}
where in the regime $|{\bf r_2}-{\bf r_1}|\gg\lambda$ we have neglected processes in
which a photon emitted into a mode $\mu$ by one atom is reabsorbed by the other one.

A closed form for the coefficients of the dipole excitations is found by summing over the
modes of the electromagnetic field and by applying the Wigner-Weisskopf approximation as
in~\cite{Dorner}. The details of the calculation are reported in App.~\ref{App:A}. The
resulting equations take the form
\begin{subequations} \label{eqns:b} \begin{eqnarray}
\dot{b}_e^{(1)}(t) &=& -\frac{\gamma}{2} b_e^{(1)}(t)
-\kappa \frac{\gamma}{2} e^{{\rm i} \omega_0 \tau } b_e^{(2)} (t-\tau) \Theta (t-\tau )\,,\nonumber\\
 \label{eqn:b1} \\
\dot{b}_e^{(2)}(t) &=& -\frac{\gamma}{2} b_e^{(2)}(t) -\kappa \frac{\gamma}{2} e^{{\rm i}
\omega_0 \tau } b_e^{(1)} (t-\tau) \Theta (t-\tau ).\nonumber\\ \label{eqn:b2}
\end{eqnarray} \end{subequations}
Equations~(\ref{eqn:b1})-(\ref{eqn:b2}) show different behaviour depending on whether
$t\le \tau$ or $t>\tau$. For $t\le \tau$ these equations are decoupled and describe
exponential damping at rate $\gamma$ of the single-atom excited-state occupation, as in
free space. After the time $\tau$, the coupling by light scattering from each atom onto
the other appears, its strength being set by the parameter $\kappa$.

We proceed by solving Eqs.~(\ref{eqn:b1})-(\ref{eqn:b2}) for an arbitrary initial state
with a single atomic excitation,
\begin{equation}\label{eqn:AnfangWinkel} \ket{\psi (0)}= \alpha_1
\ket{e,g,0}+ \alpha_2 \ket{g,e,0}\,.
\end{equation}
A simple solution is then found by using the decomposition into symmetric and
antisymmetric coefficients $C_{\pm}(t)$,
\begin{eqnarray} \label{cdup}
&&b_e^{(1)}(t)=(C_+(t)+C_-(t))/2\,,\\
&&b_e^{(2)}(t)=(C_+(t)-C_-(t))/2\,, \end{eqnarray}
obeying the differential equations \begin{equation}
\dot{C}_{\pm}(t)=-\frac{\gamma}{2} C_{\pm}(t)\mp \kappa
\frac{\gamma}{2} {\rm e}^{{\rm i} \omega_0 \tau } C_{\pm}(t-\tau )
\Theta (t-\tau )\,,
\end{equation}
whose solution is~\cite{Dung} $$ C_{\pm}(t)=C_{\pm}(0) \sum_{k=0}^{\infty} (\pm 1)^k
I_k(t),$$ with
\begin{equation} \label{J-k} I_k(t)=\frac{(- \kappa \gamm {\rm
e}^{{\rm i} \omega_0\tau })^k}{k!} (t-k\tau )^k
e^{-\frac{\gamma}{2} (t-k \tau)} \Theta (t-k\tau)\,.
\end{equation}
Correspondingly, the probability amplitudes for the excited states are
\begin{subequations} \label{eqn:excitedOhneLaser}
\begin{eqnarray}
b_e^{(1)}(t) &=& \alpha_1 \sum_k I_{2k}(t) + \alpha_2 \sum_k I_{2k+1}(t)\,, \\
b_e^{(2)}(t) &=& \alpha_1 \sum_k I_{2k+1}(t) + \alpha_2 \sum_k I_{2k}(t)\,,
\end{eqnarray}
\end{subequations}
while the probability amplitudes $b_g^{(j,\mu)}(t)$ for the emission of a photon into
mode $\mu$ by atom $j$ are given by
\begin{eqnarray} \label{eqn:MoLfieldcoeff:1}
b_g^{(1,\mu)}(t) &=& \frac{g_{\mu} {\rm e}^{-{\rm i}{\bf k_{\mu}}\cdot {\bf r_1}} }{\gamm +{\rm i}\delta_{\mu}}\,, \\
 & & \times  \sum_{k=0}^{\infty} \left [ \alpha_1 H_{2k}(t,\omega_{\mu})  + \alpha_2 H_{2k+1}(t,\omega_{\mu}) \right ]  \nonumber\\
 b_g^{(2,\mu)}(t) &=& \frac{g_{\mu} {\rm e}^{-{\rm i}{\bf k_{\mu}}\cdot {\bf r_2}} }{\gamm +{\rm i} \delta_{\mu} } \label{eqn:MoLfieldcoeff:2}\\
 & & \times  \sum_{k=0}^{\infty} \left [ \alpha_1 H_{2k+1}(t,\omega_{\mu}) + \alpha_2 H_{2k}(t,\omega_{\mu}) \right ]\,,  \nonumber
\end{eqnarray}
with
\begin{equation}\label{delta:mu}
\delta_{\mu}=\omega_0-\omega_{\mu}\,.
\end{equation}
In Eqs.~(\ref{eqn:MoLfieldcoeff:1})-(\ref{eqn:MoLfieldcoeff:2}) we assumed that the
electromagnetic field is initially in the vacuum state, $b_g^{\mu}(0)=0$, and we
introduced the function
\begin{eqnarray} && H_k(t,\omega) = \frac{
(-\kappa \gamm {\rm e}^{{\rm i} \omega \tau})^k}{k!} (t_k )^k
G_k\left [({\rm i}\delta_{\mu}+\gamma/2) t_k \right ] \Theta (t_k)\,,\nonumber\\
&&\label{eqn:MHdef}
\end{eqnarray}
with $t_k=t-k\tau$ and
\begin{equation}
\label{eqn:MLDefG} G_k(s)=~{_1}F_1(k,k+1,-s)-e^{-s}\,, \end{equation} where
$_1F_1(k,k+1,-s)$ is the confluent hypergeometric function~\cite{Abramowitz}. In the
limit $\kappa \rightarrow 0$, i.e.\ when there is no coupling between the atoms,
Eqs.~(\ref{eqn:MoLfieldcoeff:1})-(\ref{eqn:MoLfieldcoeff:2}) reduce to the usual free
space decay spectrum of two independent dipoles with linewidth $\gamma$ \cite{Heitler}.

\subsection{Perturbative solution of the Schr\"odinger equation in
presence of the laser.} \label{Sec:Eqs:Laser}

We consider now the situation that the atoms are weakly driven by a laser at intensity
$\Omega$. Hence, we set $\Omega\neq 0$ in the Schr\"odinger equation and solve the
dynamics of the new Hamiltonian assuming that $V_L$ is a weak perturbation to the atomic
dynamics. We use the ansatz for the wave function in Eq.~(\ref{eqn:wavefct}), where we
denote now the probability amplitudes by $b_e^{(j)}(t), b_g^{(j,\ell)}(t),
b_g^{(\ell)}(t) \to c_e^{(j)}(t), c_g^{(j,\ell)}(t), c_g^{(\ell)}(t)$ (with $j=1,2$ and
$\ell=\mu,\rho$). Let $\ket{\psi(0)}=\ket{g_1,g_2,0}$ be the initial state. By solving
the coupled differential equations for the probability amplitudes in first order in
$\Omega$ and in the reference frame rotating at the laser frequency $\omega_L$ we find
\begin{widetext}\begin{eqnarray}
c_{e}^{(1)}(t)&=&-{\rm i}/\hbar \int_0^t {\rm d}t'{\rm e}^{{\rm i} \omega_L t}
 \bra{e_1,g_2,0}{\rm e}^{-{\rm i} H(t-t')/\hbar} V_L(t') {\rm e}^{-{\rm i} H t'/\hbar}
 \ket{g_1,g_2,0} \nonumber \\
 &=& -{\rm i}\sqrt{2}\Omega \int_0^t {\rm d}t' {\rm e}^{{\rm i} \omega_L (t-t')}
 \bra{e_1,g_2,0} \left [{\rm e}^{-{\rm i} H (t-t')/\hbar}
 \left({\rm e}^{{\rm i}{\bf k_L}\cdot {\bf r_1}}
 \ket{e_1,g_2,0} +{\rm e}^{{\rm i}{\bf k_L}\cdot {\bf r_2}}
 \ket{g_1,e_2,0}\right)/\sqrt{2}\right ]\,. \label{eqn:MLinitialcond}
\end{eqnarray} \end{widetext}
Corresponding expressions are derived for $c_{e}^{(2)}(t)$ and $c_{g}^{(j)}(t)$. The term
inside the square bracket corresponds to the time evolution of the state
$|\beta_0\rangle=({\rm e}^{{\rm i}{\bf k_L}\cdot {\bf r_1}} \ket{e_1,g_2,0} +{\rm
e}^{{\rm i}{\bf k_L}\cdot {\bf r_2}} \ket{g_1,e_2,0})/\sqrt{2}$ when there is no laser.
Hence we can write
\begin{subequations}\label{eqn:laserkoeff}
\begin{eqnarray}
&&c_{e}^{(j)}(t)= -{\rm i}\sqrt{2}\Omega \int_0^t {\rm d}t' {\rm e}^{{\rm i} \Delta t'} b_e^{(j)}(t')\,,\\
&&c_{g}^{{(\ell)}}(t) = -{\rm i}\sqrt{2}\Omega \int_0^t {\rm d}t' {\rm e}^{-{\rm i}
\Delta_{\ell} t'} b_g^{(\ell)}(t')\,,\label{eqn:laserkoeff_field}
\end{eqnarray}
\end{subequations}
where $\ell=\rho,\mu$ and we have introduced the detunings
\begin{subequations} \begin{eqnarray}
\Delta &=& \omega_0 -\omega_L\,, \\
\Delta_{\ell} &=& \omega_{\ell}-\omega_L\,.
\end{eqnarray}
\end{subequations}
The coefficients $b_e^{(j)}$ and $b_g^{(\mu)}$ are found using the solutions derived in
Sec.~\ref{Sec:Eqs:NoLaser} when the initial state is $|\beta_0\rangle$. One gets
\begin{eqnarray}\label{eqn:MLexcited}
c_{e}^{(1)}(t) &=& \frac{-{\rm i} \Omega}{\frac{\gamma}{2}+{\rm i} \Delta }{\rm e}^{{\rm i}{\bf k_L}\cdot {\bf r_1}} \\
&  \times& \sum_{k=0}^{\infty} \left[H_{2k}(t,\omega_L )+{\rm
e}^{{\rm i}\varphi_L} H_{2k+1}(t,\omega_L ) \right ]\,,  \nonumber
\end{eqnarray}
with
\begin{equation}
\label{Phase:Laser} \varphi_L={\bf k_L}\cdot ({\bf r_2}-{\bf r_1})\,.
\end{equation}
The equation for $c_{e}^{(2)}(t)$ results from Eq.~(\ref{eqn:MLexcited}) by interchanging
the indices $1 \leftrightarrow 2$. The probability amplitudes $c_{g}^{(j,\mu)}(t)$ are
found using Eq.~(\ref{eqn:MoLfieldcoeff:1}) in Eq.~(\ref{eqn:laserkoeff_field}), assuming
that initially both atoms are in the ground state and the electromagnetic field in the
vacuum state. One gets
\begin{eqnarray} 
& &c_{g}^{(1,\mu)}(t) = -{\rm i}\frac{\Omega g_{\mu}}{\gamm +{\rm i}\delta_{\mu} }
\int_0^t {\rm d}t' {\rm e}^{{\rm i} (\omega_L -\omega_{\mu} )t'}
{\rm e}^{{\rm i}({\bf k_L}-{\bf k}_{\mu})\cdot {\bf r_1}} \nonumber\\
& & \times\sum_{k=0}^{\infty}  \left [H_{2k}(t',\omega_{\mu} ) +{\rm e}^{{\rm
i}\varphi_L} H_{2k+1}(t',\omega_{\mu} ) \right ]\,,
\label{eqn:MLfeldkoeffrech}
\end{eqnarray}
with $\delta_{\mu}$ given in Eq.~(\ref{delta:mu}). The probability amplitude
$c_{g}^{(2,\mu)}(t)$ for atom 2 is obtained by swapping the superscripts $1
\leftrightarrow 2$ in Eq.~(\ref{eqn:MLfeldkoeffrech}).

\subsection{Discussion}
\label{Sec:Rates}

The probability amplitudes of the atomic excited states in absence and in presence of the
laser, given in Eqs.~(\ref{eqn:excitedOhneLaser}) and~(\ref{eqn:MLexcited}),
respectively, are the coherent sums over contributions starting at different instants of
time $\tau_k=t-k\tau$. These contributions correspond to the effect of $k$ exchanges of a
photonic excitation between the two atoms. In particular, for the case of atom 1, the
contributions at $\tau_{2k}$ correspond to an excitation which propagated to atom 2 and
back. Hence, in Eq.~(\ref{eqn:excitedOhneLaser}) this term vanishes when initially only
atom 2 is excited. Similarly, the contributions at $t=\tau_{2k+1}$ vanish when atom 2 is
initially in the ground state. Similar considerations apply for the case in which the
laser drives the atom, Eq.~(\ref{eqn:MLexcited}).

An important property of these equations is that each term of the sum has a well-defined
phase, which is an integer multiple of $\omega_0\tau$ ($\omega_L\tau$ with the laser
excitation). At the same time the contributions are damped by an exponential function at
rate $\gamma$. Consequently, the individual terms show interference if over the time
$\tau$ they do not decay appreciably. This shows in more detail how the radiative
properties of the system are determined by the parameter $\gamma \tau$, the ratio between
the delay time and the excited state lifetime. In particular, for $\gamma\tau\gg 1$
interference plays no role, and the photonic excitation is a wave packet bouncing between
the two atoms, until its intensity is damped to zero by scattering into free space. For
$\gamma\tau\ll 1$ the terms in~(\ref{J-k}) add up coherently and interfere. The effect of
the interaction hence modifies the radiative properties of the atoms, and the dynamics
are analogous to an effective dipole-dipole interaction~\cite{Dipole-Dipole}.

In this perspective, the optical set-up composed by two atoms and the lens can be
considered like a resonator, where the atoms are mirrors of low reflectivity and
reflection bandwidth $\gamma/2$, while $2\tau$ is the round-trip time. The parameter
$\gamma\tau$ hence gives the number of modes that this peculiar "two-atom cavity"
sustains: for $\gamma\tau\gg 1$ it sustains several modes and can be considered a
"multi-mode resonator". Conversely, for $\gamma\tau\ll 1$ only a single mode of radiation
is supported, and we will denote this case as "single-mode resonator".

Using this insight, we now analyze the probability amplitude and the spectrum of the
emitted photons in the external modes labeled with $\mu$. Let us first assume that the
laser is absent, and that initially atom~1 is in the excited state, i.e.\ $\alpha_1=1,
\alpha_2=0$ in Eq.~(\ref{eqn:AnfangWinkel}). From
Eqs.~(\ref{eqn:MoLfieldcoeff:1})-(\ref{eqn:MoLfieldcoeff:2}), the amplitude probability
for the state of the field reads
\begin{eqnarray}\label{eqn:Meinsexfield} b_g^{(\mu)}(t) &=&
\frac{g_{\mu} }{\gamm +{\rm i} \delta_{\mu}}\sum_{k=0}^{\infty} \\
& &\times \Bigl[{\rm e}^{-{\rm i}{\bf k_{\mu}}\cdot {\bf r_1}} H_{2k}(t,\omega_{\mu}
)+{\rm e}^{-{\rm i}{\bf k_{\mu}}\cdot {\bf r_2}} H_{2k+1}(t,\omega_{\mu} )\Bigr]\nonumber
\nonumber
\end{eqnarray}
where the two terms under the sum account for the respective contributions of the two
atoms to the emission into the field mode. The label $k$ gives the number of photon
exchanges between the two atoms before the photon is finally emitted into the external
mode~$\mu$.

In the long time limit Eq.~(\ref{eqn:Meinsexfield}) reduces to the form
\begin{equation}\label{eqn:Mbgsteadyeins} b_g^{(\mu)}(t\to\infty ) = g_{\mu} {\rm
e}^{-{\rm i}{\bf k_{\mu}}\cdot {\bf r_1}} \frac{(\gamm +{\rm i} \delta_{\mu} )-\kappa
\gamm {\rm e}^{{\rm i} \omega_{\mu}\tau (1 + \cos \vartheta )}} {(\gamm +{\rm
i}\delta_\mu )^2-(\kappa \gamm {\rm e}^{{\rm i} \omega_{\mu} \tau })^2  }\,,\nonumber
\end{equation}
where $\vartheta$ denotes the angle between the vector ${\bf r_1}-{\bf r_2}$ and the wave
vector ${\bf k_{\mu}}$ of the mode, see Fig.~\ref{fig:FullSetup}. At $\vartheta=\pi/2$, in particular, the probability
to measure a photon in mode $\mu$ is given by \begin{equation} |b_g^{(\mu)}(t\to\infty
)|^2= \frac{g_{\mu}^2}{\frac{\gamma^2}{4} [1+\kappa \cos \omega_{\mu}\tau ]^2 +
[\omega_0-\omega_{\mu}+\kappa \gamm \sin \omega_{\mu}\tau ]^2} \end{equation} showing
that the spectrum exhibits a modulation at multiples of the frequency $1/2\tau$. In the
resonator picture, the modulation peaks are at the mode frequencies of the resonator, and
$1/2\tau$ corresponds to the free spectral range. The spectral modulation will be visible
when $\gamma\tau\gg 1$, i.e.\ when the system is in the "multi-mode-resonator" regime. On
the other hand, in the "single-mode" regime $\gamma\tau\ll 1$, one will observe a change
of the radiative linewidth, which depends on the phase $\omega_0\tau$.

When the atoms are laser-driven, the probability amplitude for the excited state
occupation in the long-time limit is
\begin{eqnarray} c_e^{(1)}(t\to\infty)&=& \frac{-{\rm i}\Omega
}{\gamm +{\rm i} \Delta} {\rm e}^{{\rm i}{\bf k_L}\cdot {\bf
r_1}}\left (
   \frac{1-K{\rm e}^{{\rm i}\varphi_L}}
{1-K^2}\right )\,,\label{c:e:1:long}
\end{eqnarray}
while the probability amplitude that mode $\mu$ is occupied by one photon scattered by
atom 1 takes the form
\begin{eqnarray}
c_g^{(1,\mu)}(t)& = &2 \pi \delta^{(t)} (\Delta_{\mu})
\frac{g_{\mu} \Omega {\rm e}^{-{\rm i}
\Delta_{\mu} t/2}}{({\rm i} \gamm -\delta_{\mu})(1-K^2) }\nonumber\\
 & \times & {\rm e}^{{\rm i}({\bf k_L}-{\bf k_{\mu}})\cdot {\bf r_1}} \left ( 1 -K {\rm e}^{{\rm i}\varphi_L}\right )
+{\rm O}(1)~~. \label{eqn:MLfeldcoessmitS}
\end{eqnarray}
Here
\begin{equation}\label{eqn:MLK} K=\kappa\gamm \frac{ {\rm e}^{{\rm i} \omega_L
\tau}}{\gamm +{\rm i}\Delta}
\end{equation}
and $\delta^{(t)}(\Delta)=\frac{1}{\pi} \frac{\sin (\Delta t/2 )}{\Delta}$ is the
diffraction function~\cite{Cohen-Tannoudji}. The second term on the right-hand side gives
no contribution to the rate of emission, so that it is not explicitly reported. Its
specific form can be found from Eq.~(\ref{eqn:MLfeldkoeffrech}) and
Appendix~\ref{Sterms}. The probability amplitudes for the second atom are obtained by
swapping the indices $1\leftrightarrow 2$. The detailed derivation of these expressions
is reported in Appendix~\ref{Sterms}.

These results are discussed for various specific limits in the following sections.

\section{Radiative Properties} \label{Sec:NoLaser}

\begin{figure}[htp] \centering
\includegraphics[width=.48\textwidth]{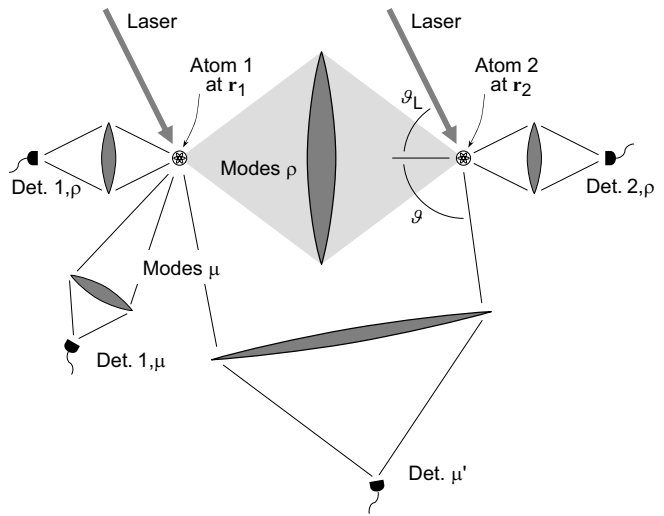}
\caption{Detailed schematic of the physical system showing the detectors which correspond
to the various measurements described in the text. Another detector $2,\mu$ would be
placed in a location equivalent to that of detector $1,\mu$, to measure the emission of
atom 2 individually.} \label{fig:FullSetup}
\end{figure}

In this section we discuss the radiative properties of the two atoms, when they are
observed together or individually, in the absence of laser excitation, and assuming that
atom 1 is initially excited. The various quantities which will be discussed correspond to
measurements with different detectors, as illustrated in the detailed set-up in
Fig.~\ref{fig:FullSetup}.

\subsection{"Multi-mode-resonator" regime}

\begin{figure}[htp] \centering
\includegraphics[width=.4\textwidth]{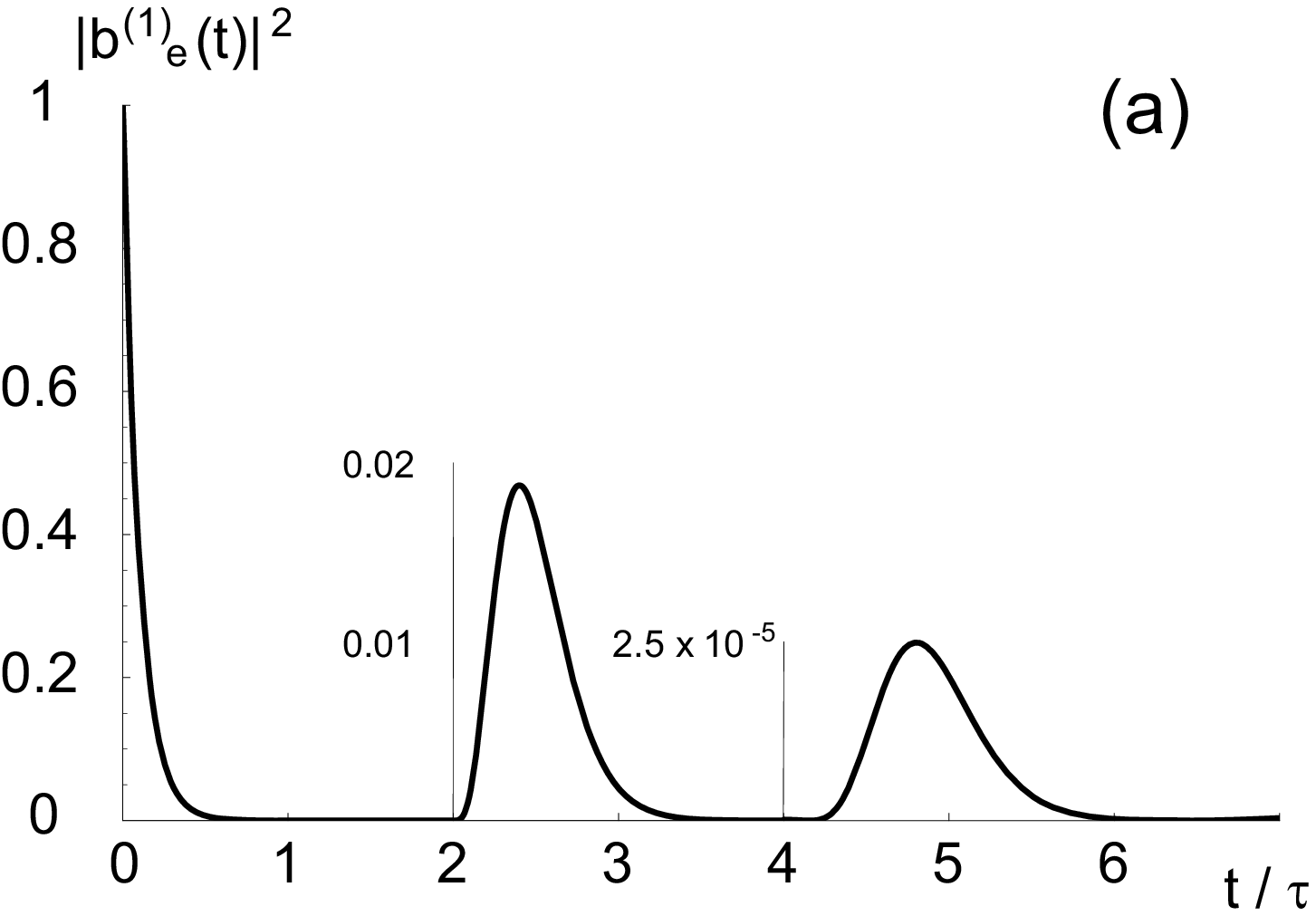}
\includegraphics[width=.4\textwidth]{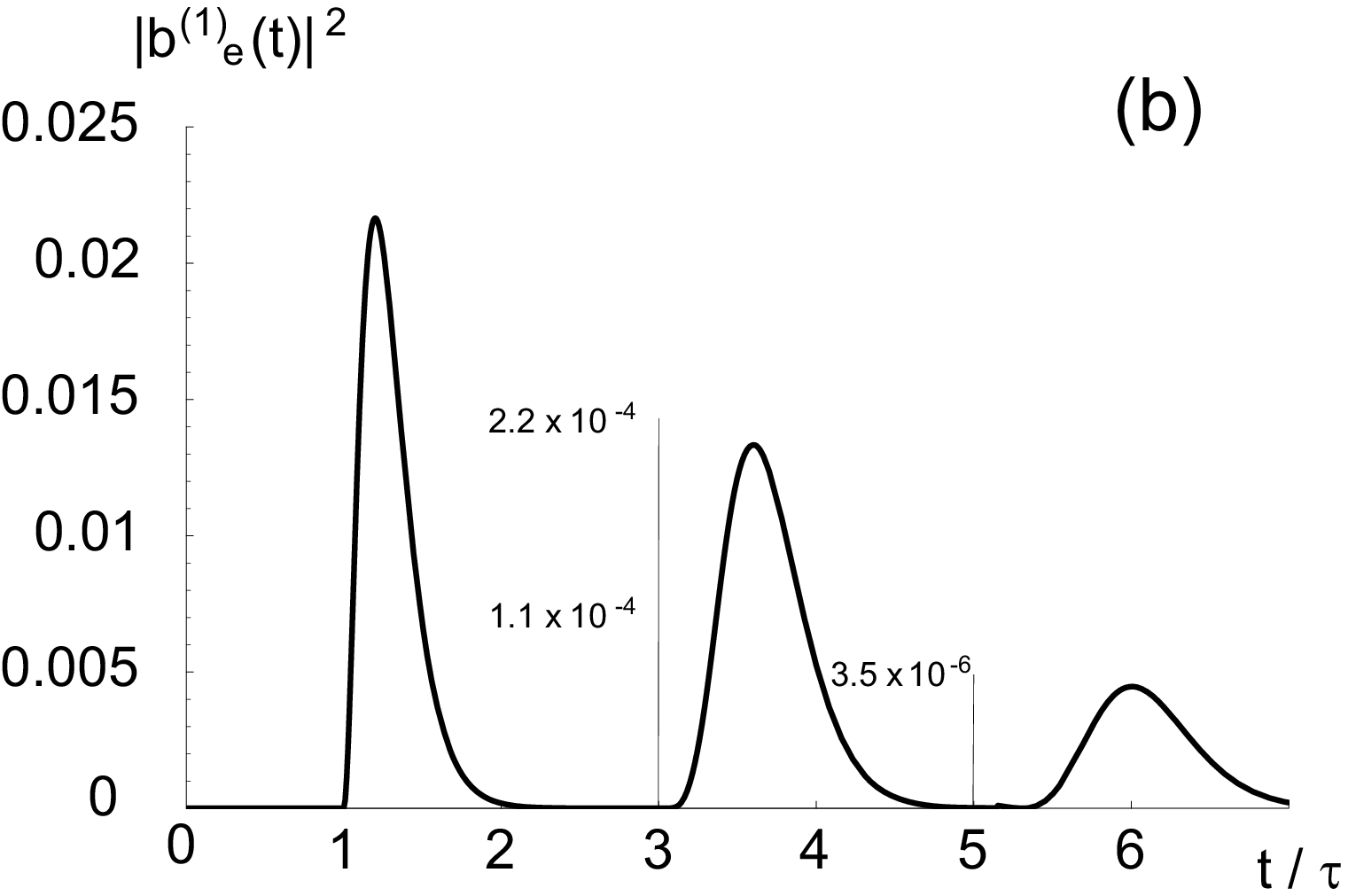}
\caption{Excited state occupation of the atoms as a function of
time in units of $\tau$, as given in
Eq.~(\ref{eqn:excitedOhneLaser}), when initially atom 1 is
excited. The other parameters are $\gamma\tau=10$, $\kappa =0.4$,
and $\omega_0 \tau=n\pi$. Note the change of vertical scale from
each maximum to the next one.} \label{fig:MExcited} \end{figure}

\begin{figure}[htp] \centering
           \includegraphics[width=.4\textwidth]
                {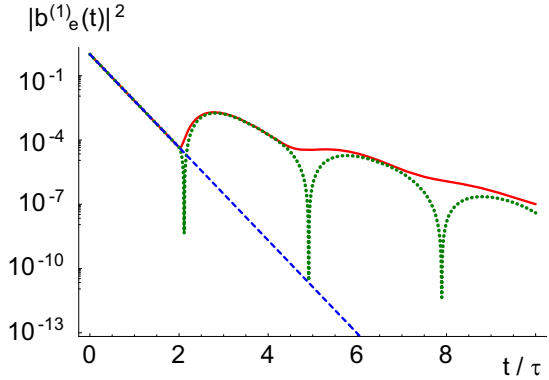}
\caption{(Color online) Logarithmic plot of the excited state occupation of atom $1$  as
a function of time in units of $\tau$, as given from Eq.~(\ref{eqn:excitedOhneLaser}).
The curves correspond to the values $\omega_0 \tau =n \pi$ (red solid line) and $\omega_0
\tau = (2n+1/2)\pi$ (green dotted line) for $\kappa=0.4$ and $\gamma \tau =5$. The dashed
line corresponds to $\kappa=0$ and is plotted for reference. } \label{fig:Log}
\end{figure}

When $\gamma\tau\gg 1$, then the transient dynamics of the system is characterized by the
two atoms exchanging a photonic excitation well localized in time. The excitation
probabilities $P_j=|b_e^{(j)}(t)|^2$, with $b_e^{(j)}(t)$ given by Eqs.~(\ref{eqn:excitedOhneLaser}),
are displayed in Fig.~\ref{fig:MExcited} as a function of time. One clearly sees that a
photonic excitation propagates back and forth between the atoms, while its amplitude is
damped due to the scattering into the external modes of the electromagnetic field. The
shape of the photonic wave packet exchanged between the two atoms changes with time: with
each bounce it acquires a more symmetric and broader shape, due to the
frequency-dependent reflection by the atoms. The broadened wave packets increasingly
overlap with time, such that interference between subsequent excitations may become
visible for long times, as shown in the example of Fig.~\ref{fig:Log}.

\begin{figure}[htp] \centering
           \includegraphics[width=.4\textwidth]
                {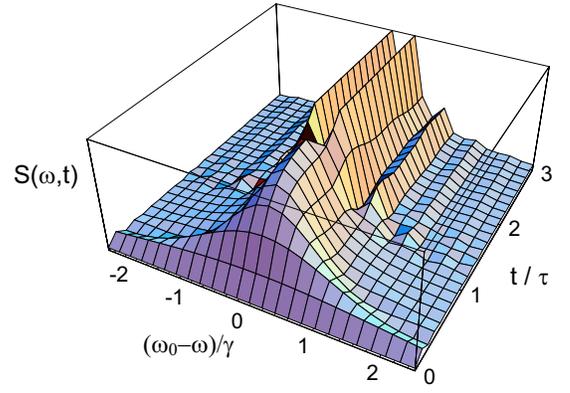}
\caption{(Color online) Probability of emission of a photon, Eq.~(\ref{Spectrum}), as a
function of frequency (in units of $\gamma$) and of time (in units of $\tau$), for the
emission angle $\vartheta =\frac{\pi}{2}$ and the phase $\omega_0 \tau = 2n\pi$.}
\label{fig:Mgroundzeit1}
\end{figure}

The effects of the coherent addition of the multiple scattering
events become more visible by inspecting the time-dependent
probability of emitting the photon into the external modes of the
electromagnetic field. In the continuum limit of
Eq.~(\ref{eqn:Meinsexfield}), it takes the
form~\cite{footnote:density} \begin{equation} \label{Spectrum}
S(\omega,t)\propto |b_g(\omega,t)|^2~~, \end{equation} which for
$t\to\infty$ coincides with the emission spectrum. An example of
$S(\omega,t)$ is shown in Fig.~\ref{fig:Mgroundzeit1}. For times
$t<\tau$, before scattering events can interfere, it exhibits a
Lorentzian form like an atom in free space, while after a time
$t>\tau$ it develops spectral modulation with peaks spaced by
$1/2\tau$.

\begin{figure}[htp] \centering
    \includegraphics[width=.4\textwidth]{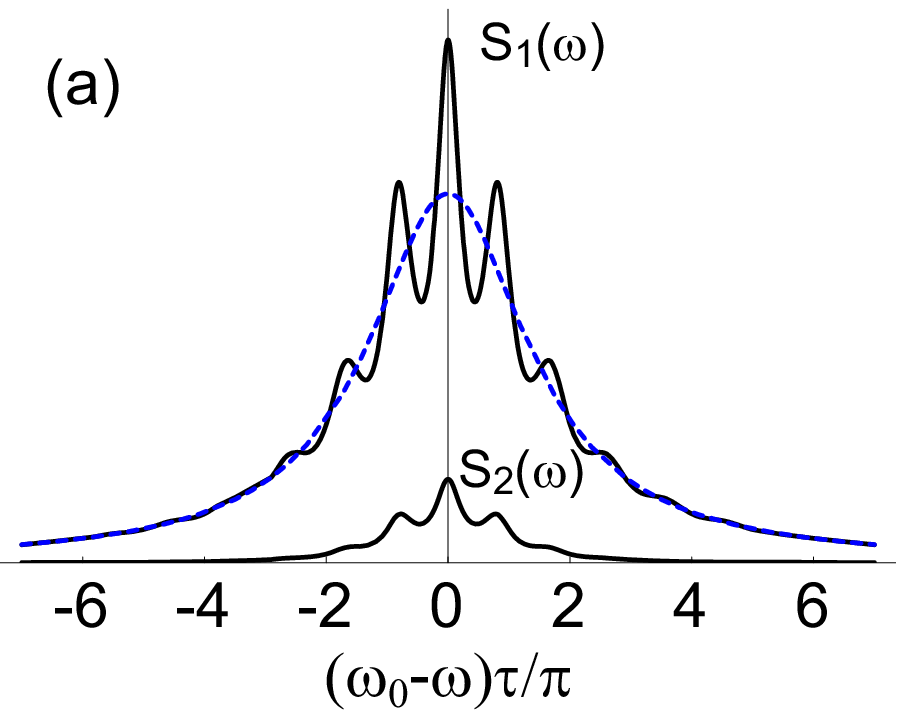}
\includegraphics[width=.4\textwidth]{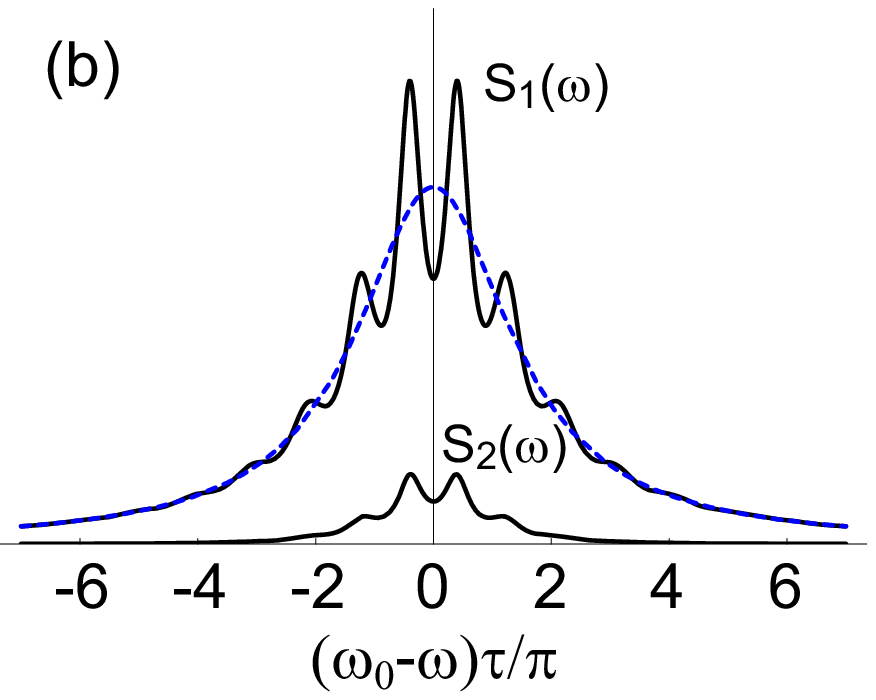}
\caption{(Color online) Spectrum of the light emitted by atom $j$,
$S_j(\omega)\propto \lim_{t\to\infty}|b^{(j)}_g(\omega,t)|^2$, as
a function of the frequency (in units of $\pi/\tau$) for (a) $\omega_0 \tau=n\pi$ and (b) $\omega_0
\tau = (2n+1) \frac{\pi}{2}$. The parameters are $\kappa =0.4$ and
$\gamma\tau= 10$. The dashed blue line gives the spectrum of the
atom when $\kappa=0$ and is plotted for comparison.}
\label{fig:spektat1} \end{figure}

The effect of the distance between the atoms on their individual emission spectra is
displayed in Figs.~\ref{fig:spektat1}(a) and~(b). In particular, the maxima of the
spectra, spaced by the "free spectral range" $1/2\tau$, shift according the optical
distance between the atoms. The visibility of modulation is larger, the closer $\kappa$
is to unity.

\subsection{"Single-mode-resonator" regime}

\begin{figure}[htp]
     \centering
     \includegraphics[width=.4\textwidth]{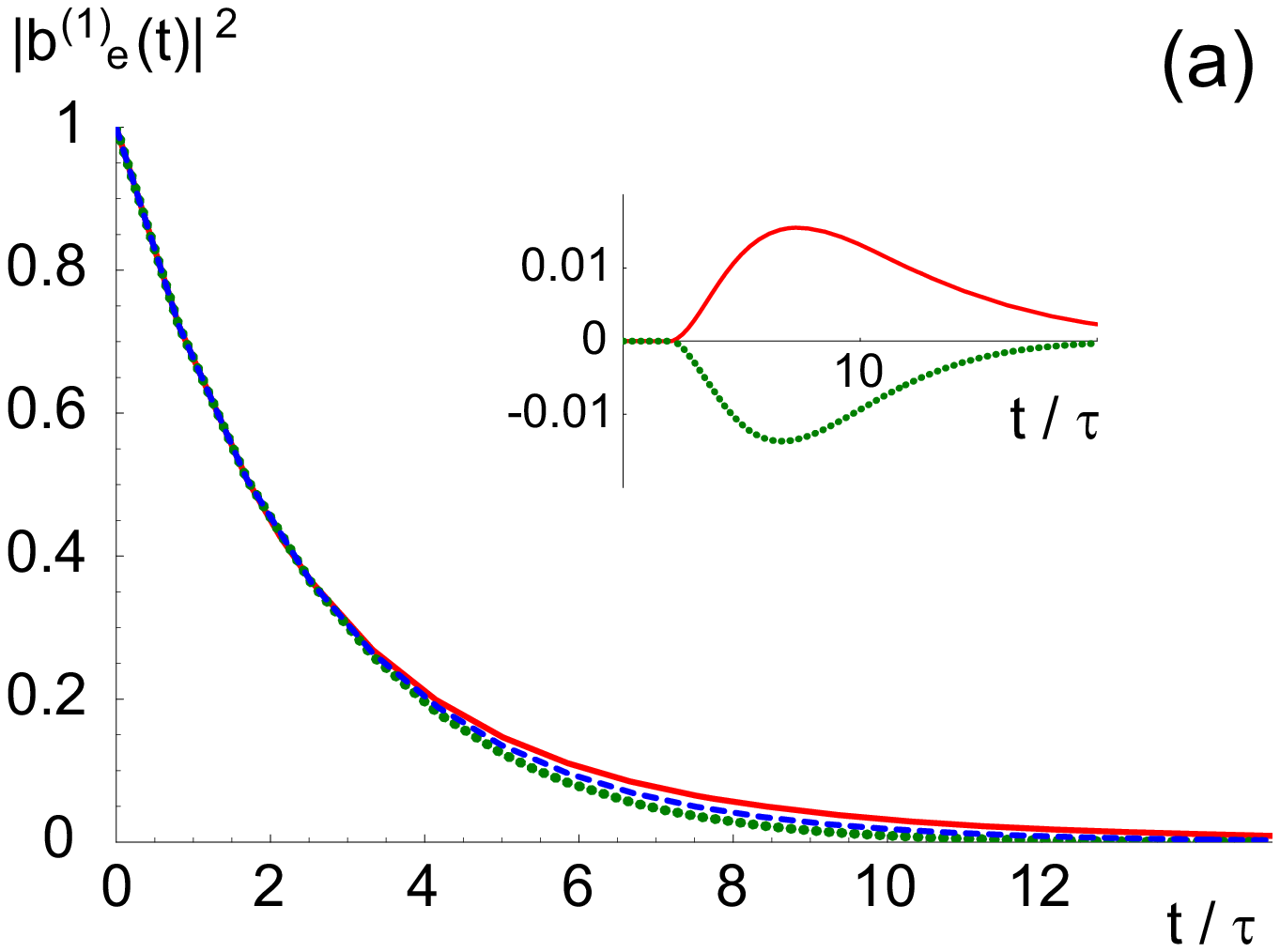}
     \hspace{.1in}
     \includegraphics[width=.4\textwidth]{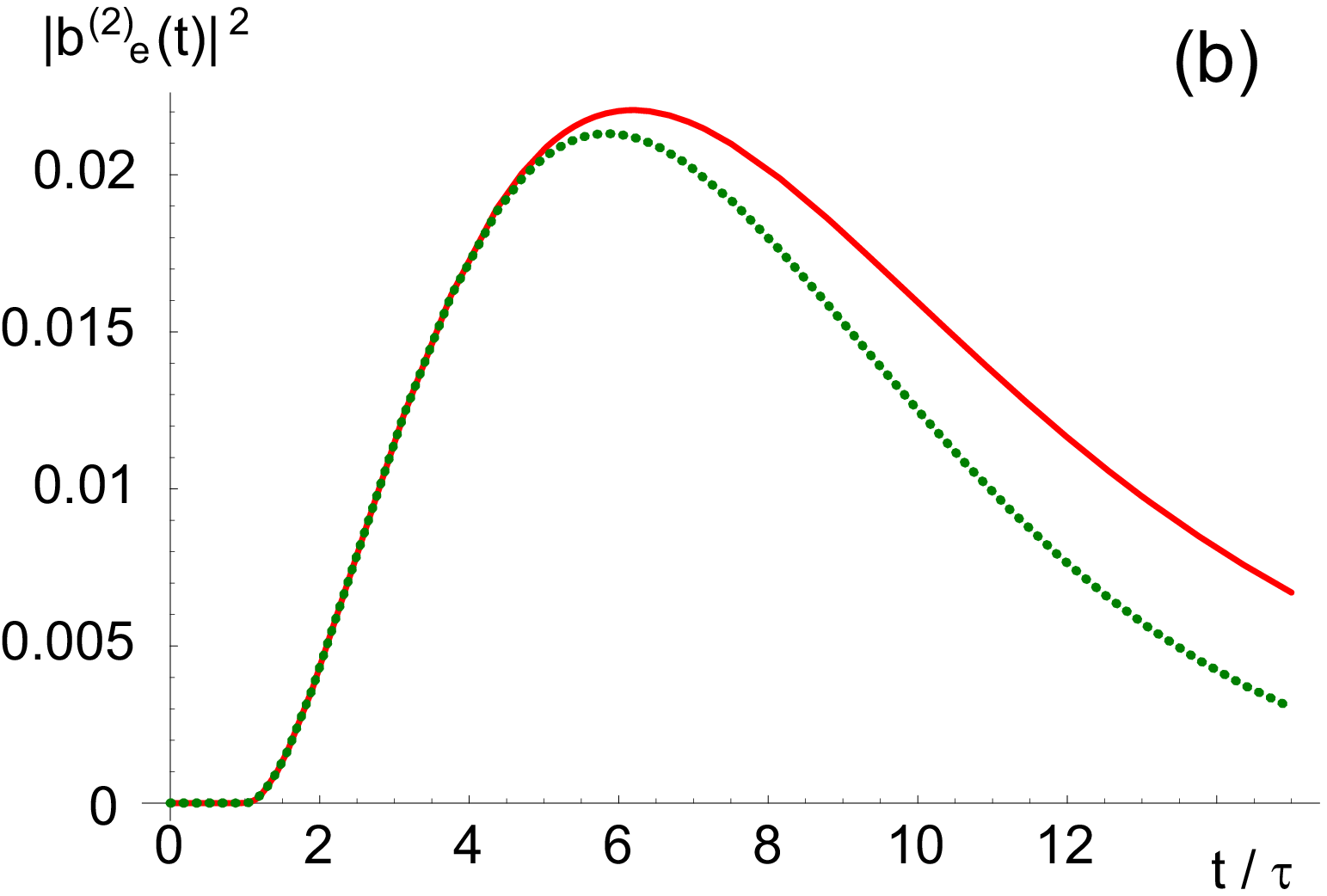}
\caption{(Color online) Excited state occupation of (a) atom 1 and
(b) atom 2 as a function of time (in units of $\tau$), as given in
Eq.~(\ref{eqn:excitedOhneLaser}), when initially atom 1 is
excited, and for $\omega_0 \tau =n\pi$ (red solid line), $\omega_0
\tau=(2n+1)\frac{\pi}{2}$ (green dotted line). The dashed blue
line is the solution for $\kappa =0$) and is displayed for
reference. The other parameters are $\gamma\tau=0.4$, $\kappa
=0.4$. The red-solid and green-dotted lines in the inset of (a)
display the difference between the value of $|b_e^{(1)}(t)|^2$ at
$\omega_0 \tau =n\pi$ and at $\omega_0 \tau=(2n+1)\frac{\pi}{2}$,
respectively, from the corresponding excited state occupation at
$\kappa=0$ as a function of time.}
\label{fig:MExcitednah} \end{figure}

\begin{figure}[htp]
     \centering
     \includegraphics[width=.4\textwidth]{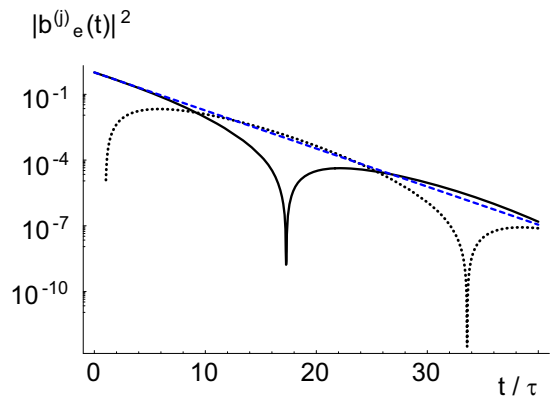}
\caption{(Color online) Logarithmic plot of the excited state occupation of atom $1$
(solid line) and atom $2$ (dotted line) as a function of time (in units of $\tau$) for
the initial state $\ket{\Psi(0)}=\ket{e,g,0}$. The other parameters are $\gamma \tau =
0.4$ , $\kappa=0.4$ and $\omega_0 \tau = (2n+1/2)\pi$. The blue dashed line shows the
corresponding atomic excitation for $\kappa=0$ and is plotted for reference.}
\label{Fig:Log:2}
\end{figure}

We now analyze the regime $\gamma\tau\ll 1$, in which several photon excitations are
exchanged between the atoms during the natural lifetime of the excited state. In
Fig.~\ref{fig:MExcitednah} the excited state populations of both atoms are displayed. As
atom 1 is initially excited, atom 2 stays in the ground state until the instant $t=\tau$,
after which its excited state occupation increases due to the interaction with the
radiation from atom 1, see Fig.\ref{fig:MExcitednah}(b). The excitation of atom 1 is
damped like in free space until time $t=2\tau$, after which the damping rate is
attenuated or enhanced depending on the relative interatomic distance, i.e.\ on
$\omega_0\tau$. The effect of the relative phase between the multiple absorption-emission
events is more evident when plotting the excitation probabilities on a logarithmic scale,
as shown in Fig.~\ref{Fig:Log:2}.

\begin{figure}[htp]
     \centering
     \includegraphics[width=.4\textwidth]{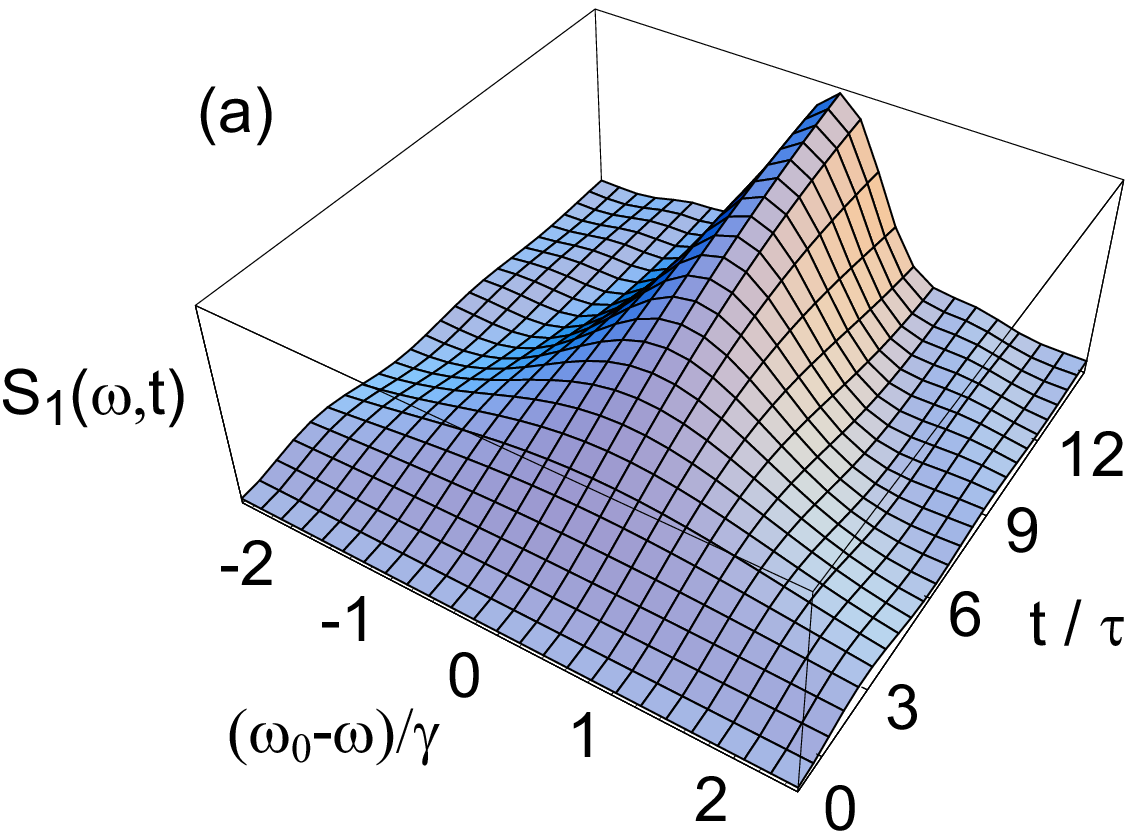}
     \hspace{.1in}
     \includegraphics[width=.45\textwidth]{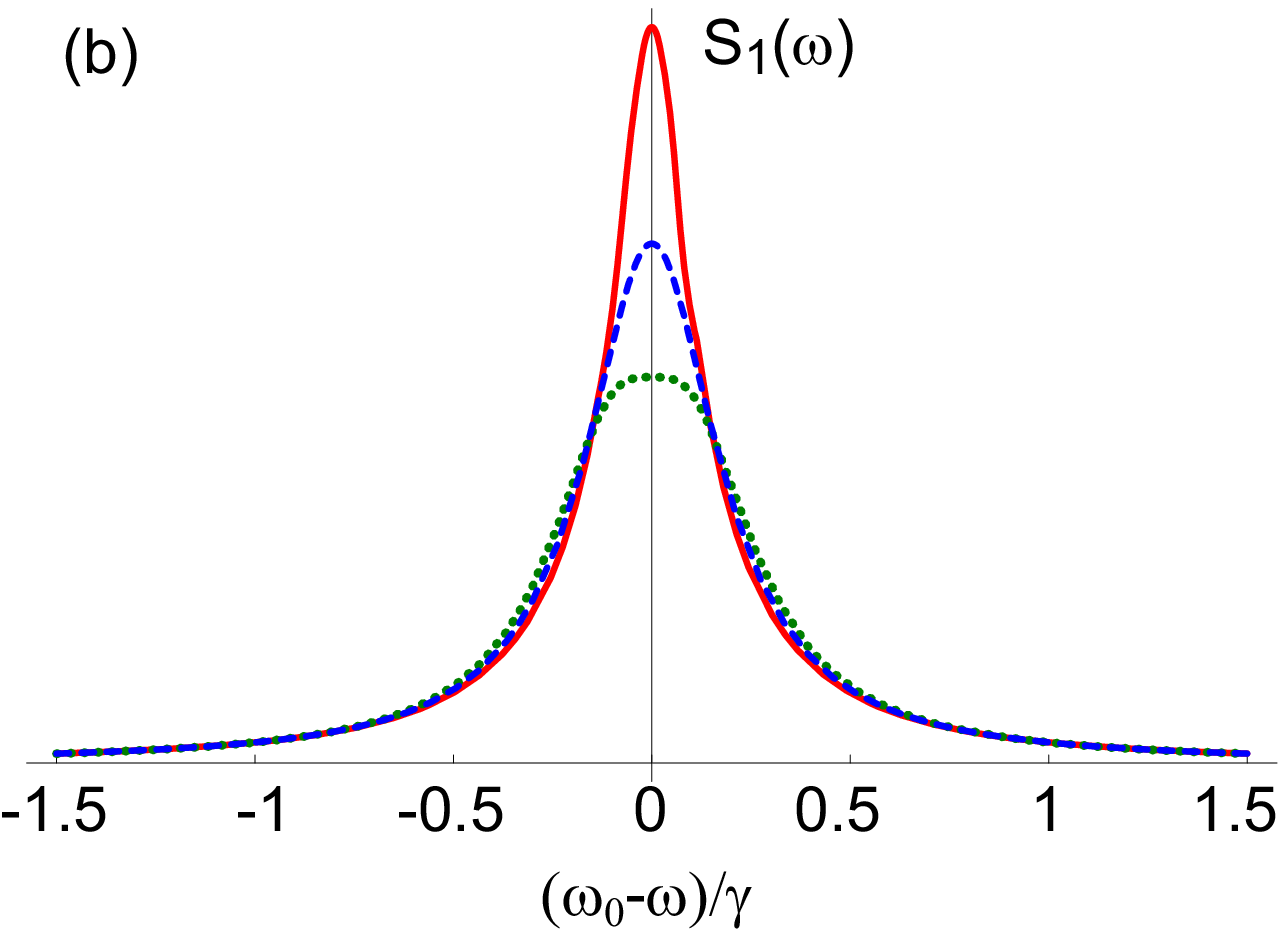}
\caption{(Color online) (a) Probability of emission of a photon by atom 1,
$S_1(\omega,t)\propto |b_g^{(1)}(\omega,t)|^2$, as a function of the frequency of the
emitted photon (in units of $\gamma$) and of time (in units of $\tau$) for $\omega_0 \tau
= n\pi$. (b) Spectrum of emission for atom 1, $S_1(\omega)$, for $\omega_0 \tau =n\pi $
(red solid line) and $\omega_0 \tau = (2n+1)\frac{\pi}{2}$ (green solid line). The other
parameters are $\gamma\tau =0.4$ and $\kappa =0.4$. The dashed blue line gives the
emission spectrum in free space and is plotted for comparison. } \label{fig:spektrum1}
\end{figure}

Figure~\ref{fig:spektrum1}(a) displays the emission probability $S(\omega,t)$ as a
function of frequency and time, showing that it is always a single-peaked curve, whose
width varies with time. Figure~\ref{fig:spektrum1}(b) displays the emission spectrum of
the first atom in comparison with the one in free space for different values of the
parameter $\omega_0\tau$, showing that depending on the relative distance one can observe
subradiant or superradiant emission. The atomic interaction is hence a retarded
dipole-dipole interaction, mediated by the photonic excitation over the interatomic
distance.

\subsection{Two atoms vs.\ single atom}

The cases studied so far share several analogies with the radiative properties of a
single atom in front of a mirror, analyzed for instance in~\cite{Alber,Dorner}. In
particular, in~\cite{Alber} Alber studied the dynamics of one photon coupled to one atom
at the center of a spherically symmetric cavity with perfect reflectivity. Depending on
the radius of the cavity mirror, and thus on the time the photon needs to travel to the
mirror and back in relation to the atomic decay time, Alber defines the small- and
large-cavity limit, whereby in the first case the atom-cavity system is characterized by
a delocalized excitation, while in the second case a photonic wave packet propagates back
and forth exciting periodically the atom. Although our system is a low-quality resonator,
the multi-mode cavity that the two atoms form for $\gamma\tau\gg 1$ is analogous to the
large-cavity limit in~\cite{Alber}.

\begin{figure} [htp] 
     \centering
     \includegraphics[width=.35\textwidth]{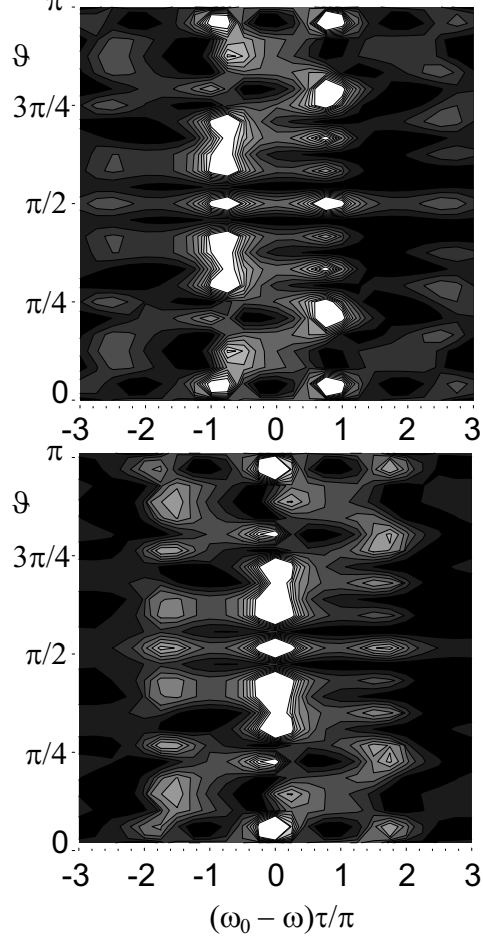}
\caption{Contour plots of the emission spectrum from both atoms
$S(\omega)$, Eq.~(\ref{Spectrum}), as a function of the frequency
$\omega$ (in units of $\pi/\tau$) and of the angle of emission
$\vartheta$, for the initial state $\ket{\Psi
(0)}=\frac{1}{\sqrt{2}}( \ket{e,g,0}+\ket{g,e,0})$. The other
parameters are $\gamma \tau=10$, $\kappa =0.4$, and $\omega_0 \tau
=2n \pi$ (top panel), $\omega_0 \tau =(2n +1)\pi$ (bottom panel).
} \label{fig:Mgsymcontour} \end{figure}

A very close analogy exists between our system and the system discussed in~\cite{Dorner},
where Dorner and Zoller investigated the case of an atom interacting with its own light
back-reflected by a distant mirror~\cite{Dorner}. In particular, the dynamics of two
atoms exchanging photons via the lens share strong analogies with the one of an atom
interacting with its mirror image, if one restricts the Hilbert space to only one
excitation, and if the atoms are initially prepared in a symmetric state with
$\alpha_1=\alpha_2=1/\sqrt{2}$ in Eq.~(\ref{eqn:AnfangWinkel}). For this initial state,
the time evolution of the excitation of one of the atoms is the same as the one of the
atom in front of the mirror. The probability amplitude for photon emission, however,
shows some differences between the two cases. Figure~\ref{fig:Mgsymcontour} displays the
emission spectrum as a function of the emission angle and of the frequency. Here, the
oscillation of the intensity as a function of the angle of emission $\vartheta$ is indeed
a exclusive property of the two-atom case, arising from the fact that the light emitted
from the two scatterers interferes in the far field.

 \section{Light scattering}
\label{Sec:Laser}

\begin{widetext}

 \begin{figure}[htp]     \centering
          \includegraphics[width=.4\textwidth]{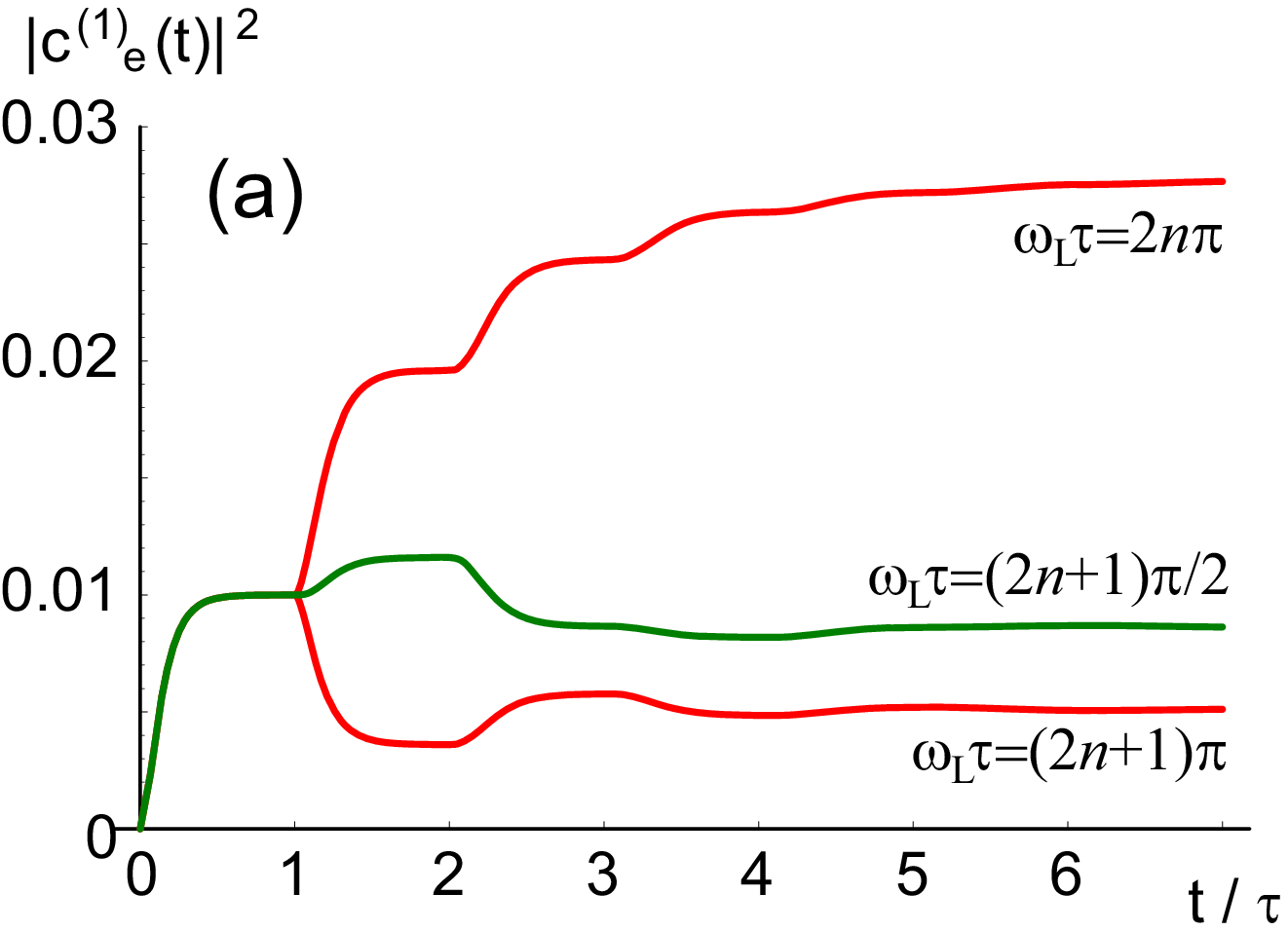}
          \includegraphics[width=.4\textwidth]{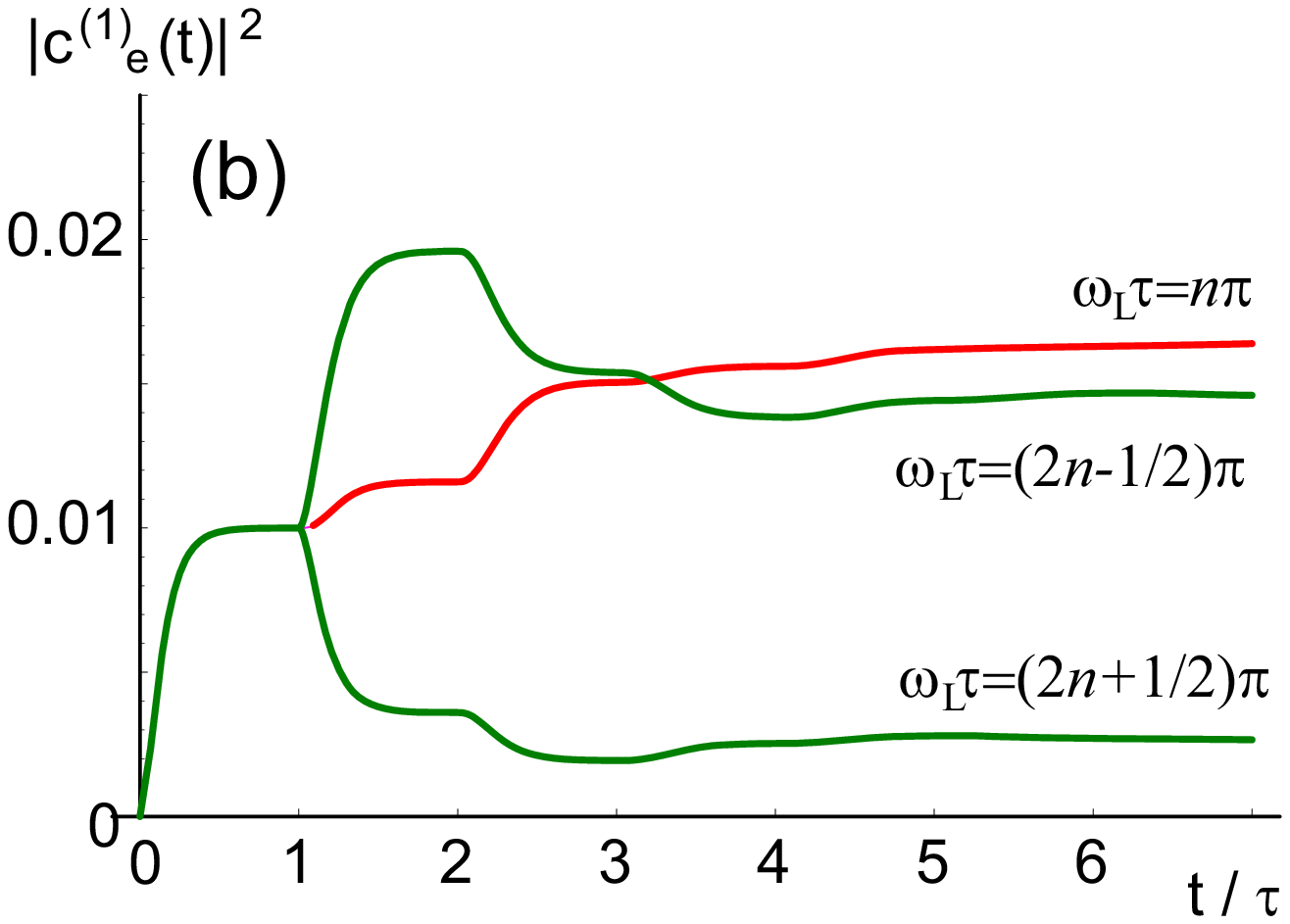}
        \includegraphics[width=.4\textwidth]{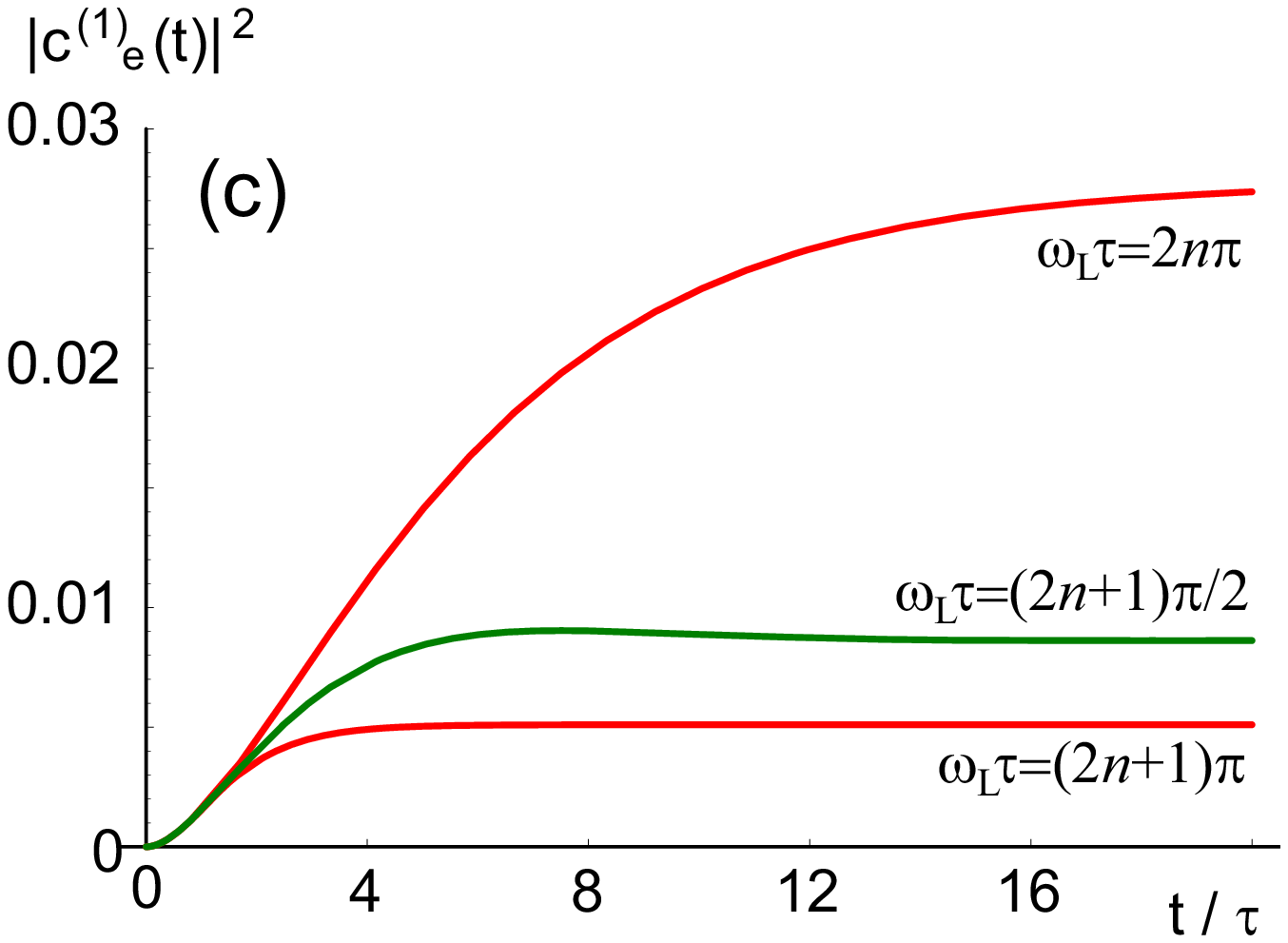}
         \includegraphics[width=.4\textwidth]{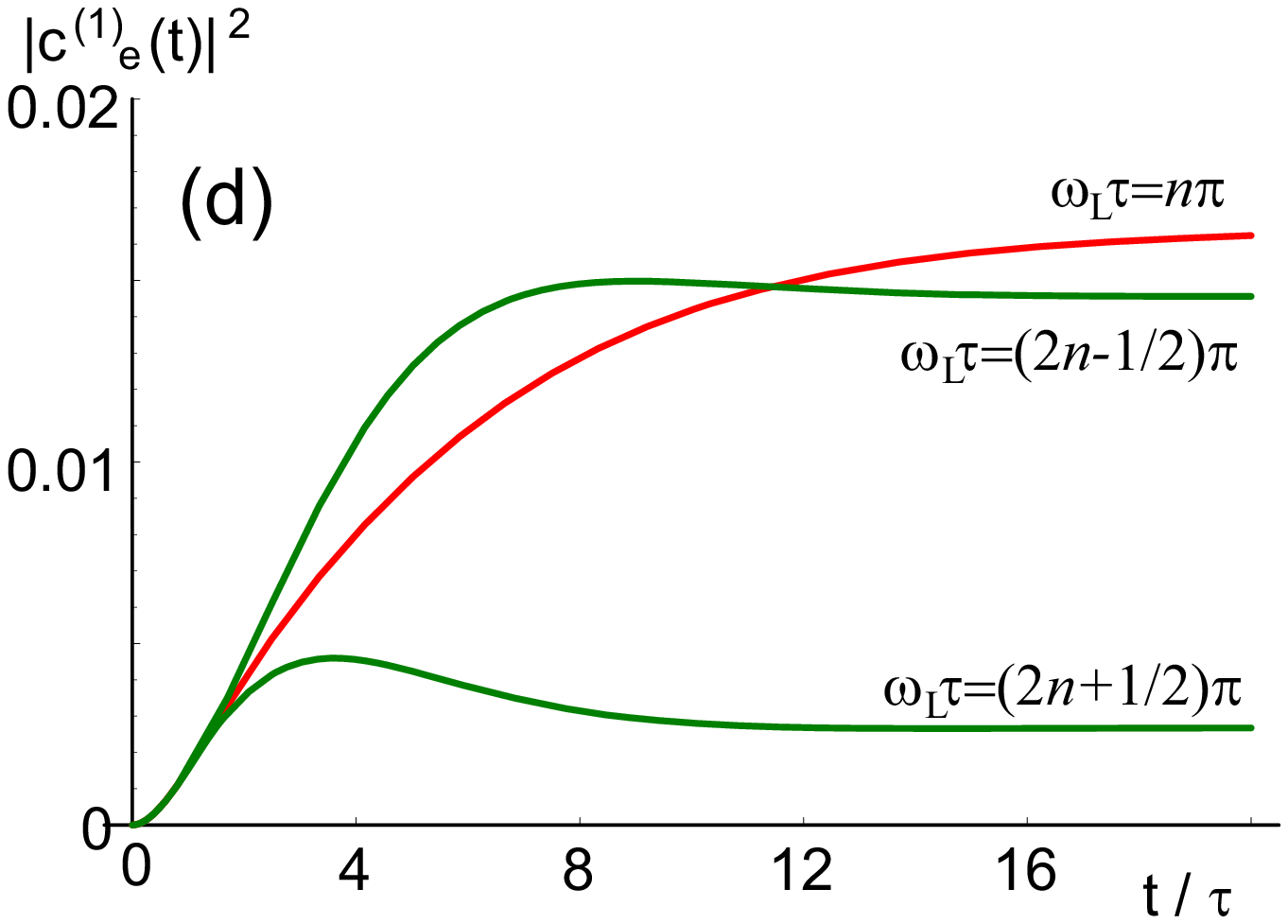}
\caption{(Color online) Excited state population of atom $1$ when both atoms are driven
by the laser, as evaluated from Eq.~(\ref{eqn:MLexcited}), as a function time (in units
of $\tau$). The figures are evaluated for $\kappa=0.4$, $\Omega=0.05\gamma$, $\Delta=0$,
and for $\gamma\tau=20$ (upper row) and for $\gamma \tau =1$ (lower row). The subplots
(a) and~(c) refer to the case $\varphi_L=(2n+1)\pi$, (b) and~(d) to the case
$\varphi_L=(2n-1/2)\pi$. The value of the phase $\omega_L\tau$ for each curve is
explicitly given in the plots.} \label{fig:MLExcited}
\end{figure}
\end{widetext}

In this section we analyze the scattering properties of the system when the atoms are
driven by a laser below saturation. In this case the time evolution of the excited state amplitudes,
Eqs.~(\ref{eqn:MLexcited}), describes the photon exchange between the two atoms, which
now additionally interferes with the incident laser light. For $\gamma\tau\gg 1$,
step-wise dynamics with the characteristic time step $\tau$ are visible in the excited
state occupation of each atom, as displayed in Fig.~\ref{fig:MLExcited}(a) and~(b). For
different distances between the atoms, and hence different phases of the various
contributions, the discontinuities in the curves at multiples of $\tau$ show constructive
or destructive interference, while for long times $t\gg \tau$ the excited-state
population tends to a steady state value. In the limit $\gamma\tau\le 1$, displayed in
Fig.~\ref{fig:MLExcited}(c) and~(d), the curves are smooth and tend to the same steady
state values. This stationary value depends on the two phases $\varphi_L$ (the laser
direction) and $\omega_L\tau$ (the optical path length between the atoms), according to
\begin{equation}\label{eqn:MLexcsteady}
|c_{e}^{(1)}(t\to \infty )|^2 = \frac{\Omega^2}{(\frac{\gamma^2}{4}+\Delta^2) |1-K^2|^2}
\left | 1-K {\rm e}^{{\rm i}\varphi_L}\right |^2\,,
\end{equation}
where $K$ is given in Eq.~(\ref{eqn:MLK}) and is proportional to the coupling strength
$\kappa$ between the two ions. Eq.~(\ref{eqn:MLexcsteady}) does not depend on the
parameter $\gamma\tau$, which affects only the transient dynamics. When the atoms are not
coupled, $\kappa=0$, one recovers the free-space steady state value, as found for an atom
which is driven by a weak laser~\cite{Cohen-Tannoudji}.

We note that for the specific value $\varphi_L=(2n+1)\pi$ we
obtain \begin{equation} |c_{e}^{(1)}(\infty )|^2 =
\frac{\Omega^2}{\tilde{\gamma}_L^2/4+\tilde{\Delta}^2}\,,
\end{equation} which is the free-space formula with modified decay
rate and detuning, \begin{eqnarray}
\tilde{\gamma} &=& \gamma \, (1-\kappa \cos \omega_L \tau )\,, \nonumber \\
\tilde{\Delta} &=& \Delta -\kappa \gamm \sin \omega_L \tau\,.
\nonumber
\end{eqnarray}
This result coincides with the excited state population of a single atom subject to
interference between the laser excitation and the light back-scattered from a
mirror~\cite{Dorner}. This equivalence holds only for the particular value
$\varphi_L=(2n+1)\pi$ but not in the general case.

In order to get some more insight, we analyze Eq.~(\ref{eqn:MLexcsteady}) for $\kappa\ll
1$. At first order in $\kappa$ the excited state population of the first atom takes the
value
\begin{eqnarray}
\label{eqn:MLcexsteadyphi}
|c_{e}^{(1)}(t\to \infty )|^2 &\approx & \frac{\Omega^2}{\gamma^2/4+ \Delta^2} \nonumber \\
& \times & \left (1+2 \kappa A \cos
\left[ \omega_L \tau +\varphi_L-\phi \right ]\right )\,,\nonumber \\
\end{eqnarray}
with
\begin{eqnarray} \label{eqn:AundPhi}
&&A = \sqrt{\frac{\gamma^2/4}{\gamma^2 /4+\Delta^2}}\,, \nonumber \\
&&\tan \phi= \frac{2\Delta}{\gamma}\,.
\end{eqnarray}
The result for atom 2 is found from Eq.~(\ref{eqn:MLcexsteadyphi}) by swapping the
subscripts $1 \leftrightarrow 2$, i.e.\ by changing the sign of $\varphi_L$.
Equation~(\ref{eqn:MLcexsteadyphi}) shows how the excited state population is enhanced or
suppressed as the parameter $\omega_L\tau$ is changed. This change in the atomic
spontaneous emission rate, as well as a shift of the atomic resonance frequency, both
controlled by the parameter $\omega_L\tau$, are manifestations of the modification of the
radiative properties of the atoms due to their mutual interaction. Analogous frequency
shifts in a single atom interacting with itself via a mirror have been experimentally
observed by Wilson et al.~\cite{Wilson}.

\subsection{Intensity of the scattered light}

Let us now consider the intensity of the light scattered by the laser-driven atoms, for
several of the measurement set-ups illustrated in Fig.~\ref{fig:FullSetup}.

\begin{figure}[htp] \centering
\includegraphics[width=.4\textwidth]{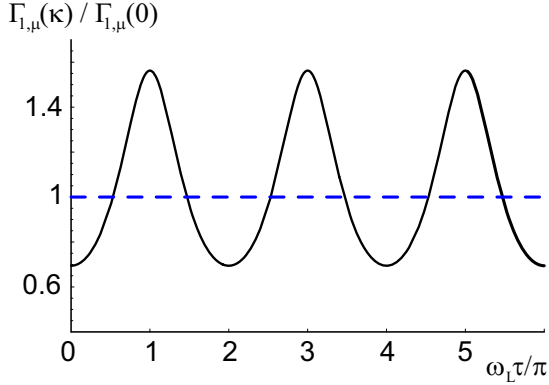}
\caption{(Color online) Intensity of the light emitted by one atom
into free space, as a function of $2\omega_L \tau$ and for
$\kappa=0.2$, when the system is driven by a laser at frequency
$\omega_L=\omega_0$ and at $\varphi_L=0$. The intensity is
normalized to the value obtained without coupling, $\kappa=0$
(blue dashed line). } \label{fig:Multiple} \end{figure}

First we consider the situation that the detection apparatus resolves the atomic
positions and therefore sums up incoherently the photons emitted by the atoms (detector
$1,\mu$ plus detector $2,\mu$). Then the detection rate is
$\Gamma_{\mu}=\Gamma_{1,\mu}+\Gamma_{2,\mu}$, whereby $\Gamma_{j,\mu}= \lim_{t \to
\infty} |c_{g}^{(j,\mu)}(t)|^2/t$. Using Eq.~(\ref{eqn:MLfeldkoeffrech}) we find
\begin{eqnarray} \label{eqn:Gamma:inc}
\Gamma_{1,\mu}=\Gamma_{2,\mu}=2\pi \delta (\omega_{\mu}-\omega_L )\frac{g_{\mu}^2
\Omega^2}{(\frac{\gamma^2}{4}+\delta_{\mu}^2 )} \frac{\left |1-K{\rm e}^{{\rm
i}\varphi_L}\right |^2}{|1-K^2|^2}\,,
\end{eqnarray}
where $K$ is given in Eq.~(\ref{eqn:MLK}). For $K=0$, i.e.\ in absence of the optical
element coupling the two atoms, the signal reproduces the free-space resonance curve of
the atomic dipole. For $K\neq 0$ it shows two modulations, with the phase $2\omega_L\tau$
(through $K^2$ in the denominator) and with the phase $\omega_L\tau+\varphi_L$ (in the
numerator). The first one corresponds to previously emitted light returning to the same
atom after scattering from the other one, the other modulation is produced by scattered
laser light arriving from the other atom. In general, the modulations show how the
scattering of a single atom is modified by the presence of another identical scatterer at
a fixed distance. The maximum enhancement, when all scattering terms add up coherently,
is found for $2\omega_L\tau=2\pi$, $\varphi_L=0$ and $\Delta=0$, and is equal to
$(1+\kappa)/(1-\kappa)$. For $\kappa=0.2$ it gives an enhancement of the signal of the
order to 150\%, as displayed in Fig.~\ref{fig:Multiple}. For the case of a single atom
interacting with itself via a distant mirror, analogous signals have been experimentally
observed in Refs.~\cite{Eschner2001, Wilson}.

When the light emitted by the two atoms is superposed coherently on a detector (labeled
$\mu'$ in Fig.~\ref{fig:FullSetup}), the system is analogous to a double-slit
set-up~\cite{Itano} with the important difference that the atoms additionally interact by
photon exchange via the lens. In this case, the corresponding detection rate is
$\Gamma_{\mu}^{\prime}=\lim_{t \to \infty} |c_{g}^{(\mu)}(t)|^2/t$ and takes the explicit
form
\begin{widetext}
\begin{eqnarray} \label{eqn:Gamma:coh}
\Gamma_{\mu}^\prime=8\pi\delta (\omega_{\mu}-\omega_L ) \frac{g_{\mu}^2
\Omega^2}{(\frac{\gamma^2}{4}+\delta_{\mu}^2 )|1-K^2|^2} \left |\cos[({\bf k_L}-{\bf
k}_{\mu})\cdot ({\bf r_1}-{\bf r_2})/2]-K\cos[({\bf k_L}+{\bf k}_{\mu})\cdot ({\bf
r_1}-{\bf r_2})/2]\right |^2\,.
\end{eqnarray}
\end{widetext}
For $K=0$, the observed spatial interference is the one of a double-slit set-up with two
coherently driven sources~\cite{Itano, Skornia}. When $K\neq 0$, these properties are
modified by the multiple scattering. An important special case is when the direction of
emission is ${\bf k_{\mu}}=-{\bf k_L}$, corresponding to the coherent backscattering
direction~\cite{Wickles}, where one always finds a spatial maximum of the scattered
intensity.

We now consider a set-up in which one observes the modes $\rho$,
through which the atoms interact (detectors $1,\rho$ and $2,\rho$
in Fig.~\ref{fig:FullSetup}). This corresponds to the measurement
arrangement in~\cite{Eschner2001}. In this case, the rate at
detector $1,\rho$ is given by $\Gamma_{1,\rho}=\lim_{t \to \infty}
|c_{g}^{(1,\rho)}(t)+ c_{g}^{(2,\rho')}(t)|^2/t$, superposing the
light emitted by atom 1 directly into the detector with the light
emitted by atom 2 towards atom 1, and then into the detector,
whereby ${\bf k}_{\rho}$ and ${\bf k}_{\rho}^{\prime}$ are
transformed into each other by the lens. It reads \begin{widetext}
\begin{equation} \label{eqn:Gamma:0} \Gamma_{1,\rho}=8\pi \delta
(\omega_{\rho}-\omega_L ) \frac{g_{\rho}^2
\Omega^2}{(\frac{\gamma^2}{4}+\delta_{\rho}^2 )|1-K^2|^2} \left
|\cos[(\varphi_L-\omega_L\tau)/2]-K
\cos[(\varphi_L+\omega_L\tau)/2]\right |^2 \,,\end{equation}
\end{widetext} where we used that ${\bf k}_{\rho'}\cdot {\bf
r_2}-{\bf k}_{\rho}\cdot{\bf r_1}\to \omega_L\tau$ via the optical
element. The corresponding rate $\Gamma_{2,\rho}$ is found by
changing the sign of $\varphi_L$ in Eq.~(\ref{eqn:Gamma:0}). The
total rate
$\Gamma_{\rho}^{(0))}=\Gamma_{1,\rho}^{(0)}+\Gamma_{2,\rho}^{(0)}$
was measured in Ref.~\cite{Eschner2001} in an optical set-up,
which was characterized by small values of $\kappa$.
Taking $\kappa\ll 1$, the total rate $\Gamma_{\rho}^{(0)}$ reads
\begin{widetext} \begin{eqnarray} \label{Gamma:0:tot}
\Gamma_{\rho}^{(0)} \propto 1+ \cos \varphi_L\cos \omega_L \tau-
\kappa A \left [\cos\phi+2 \cos \varphi_L \cos (\omega_L \tau
-\phi )+\cos (2\omega_L \tau -\phi ) \right ]+{\rm
O}(\kappa^2)\,,\end{eqnarray} \end{widetext} where we omitted global
constant factors, and $\phi$ and $A$ are defined in
Eq.~(\ref{eqn:AundPhi}). We observe that at zero order in $\kappa$
an interference pattern appears as a function of $\omega_L\tau$,
i.e.\ by changing the optical path between the ions. This is the
classical interference of the light elastically scattered from
both atoms into the same detector. The interference has visibility
\begin{equation} \label{contrast:1} {\mathcal V}_1=|\cos\varphi_L|\,,
\end{equation} which is maximum when $\varphi_L=n\pi$, with $n$
integer, and which vanishes when $\varphi_L=(2n+1)\pi/2$. This is
a consequence of summing the signals from the two detectors, whose
individual interference patterns may be shifted depending on
$\varphi_L$. It provides an explanation for the low contrast
interference observed in the experiment of
Ref.~\cite{Eschner2001}.

The vanishing contrast when the two signals are perfectly
anti-correlated provides a condition where the higher order
effects in $\kappa$, and thereby the interaction of the atoms, are
particularly evident. Choosing this specific condition, by varying
the optical path length between the ions one observes an
interference pattern at twice the frequency of the classical
interference, i.e.\ oscillating with $2\omega_L\tau$, with a shift
$\phi$ determined by the detuning $\Delta$, and whose visibility
is given by \begin{equation} {\mathcal V}_2=\kappa
A=\frac{\kappa}{\sqrt{1+4\Delta^2/\gamma^2}}\,, \end{equation} where
we used Eq.~(\ref{eqn:AundPhi}). The visibility is maximum at
atomic resonance. The doubled frequency of the interference
(compared to the classical one) with the interatomic distance
shows that it is caused by two partial waves originating from the
same atom, one reaching directly the detector and the other being
back-scattered once by the other atom. Analogously, processes
where the same wave is scattered $n$ times by the atoms give rise
to interference terms with frequency $n\omega_L \tau$ and at
higher order in $\kappa$.

\subsection{Intensity-Intensity Correlations}

We now study the intensity-intensity correlations in this set-up, assuming that two
detectors are placed in the far field of the scattered light at positions ${\bf x_1}$ and
${\bf x_2}$ (corresponding to two detectors $\mu'$ in Fig.~\ref{fig:FullSetup} at angles
$\vartheta_{1}$ and $\vartheta_{2}$). We denote by ${\mathcal G}^{(2)}({\bf x_1},t; {\bf
x_2},t+t')$ the (un-normalized) intensity-intensity correlation function for measuring a
photon at time $t$ and position ${\bf x_1}$, and another at ${\bf x_2}$ after an interval
$t'$. It reads~\cite{Skornia}
\begin{widetext} \begin{eqnarray} \label{Eq:G2} {\mathcal G}^{(2)}({\bf
x_1},t; {\bf x_2},t+t') = \sum_{\zeta ,\nu ,\lambda, \rho =1,2} {\rm e}^{{\rm i} k
\left(({\bf r}_{\lambda}-{\bf r}_{\nu}) \cdot {\bf \hat x_1} + ({\bf r}_{\xi}-{\bf
r}_{\zeta}) \cdot {\bf \hat x_2}\right)} ~ \Av{\sg^+_{\lambda}(t)\sg^+_{\xi}(t+t'
)\sg_{\nu}(t+t') \sg_{\zeta} (t)}\,, \end{eqnarray} \end{widetext} where $k=|{\bf k_L}|$ and
${\bf \hat x}={\bf x}/|{\bf x}|$. Assuming that the atoms are driven by the laser and
have reached the steady state, Eq.~(\ref{Eq:G2}) depends solely on the time $t'$ elapsed
between the two detection events. In this limit, we evaluate its explicit form in
perturbation theory for the atom-photon interaction, and find the expression
\begin{widetext}
\begin{eqnarray}\label{eqn:G2allg} G^{(2)}(\varphi_1,\varphi_2;t') &=& \left . \frac{16
\Omega^4 }{(\gamma^2 +4 \Delta^2 )^2}\frac{1} {|1-K^2|^2 } \right | \left ( 1+{\rm
e}^{{\rm i} (\varphi_1 +\varphi_L )} -K
({\rm e}^{{\rm i}\varphi_1} +{\rm e}^{{\rm i} \varphi_2}) \right ) \\
& \times & \left [ \left(1+{\rm e}^{{\rm i} (\varphi_2 +\varphi_L
)}\right) \sum_{k} H_{2k}(t',\omega_L) +({\rm e}^{{\rm i}
\varphi_L }+{\rm e}^{{\rm i} \varphi_2} )
\sum_{k} H_{2k+1}(t',\omega_L) \right ] \nonumber \\
& + &  \left . {\rm e}^{{\rm i} (\varphi_L -\Delta t')} (1-K \cos
\varphi_L  ) \left[({\rm e}^{{\rm i} \varphi_1}+{\rm e}^{{\rm i}
\varphi_2} ) \sum_k I_{2k} + \left(1+{\rm e}^{{\rm i}(\varphi_1
+\varphi_2 )}\right) \sum_k I_{2k+1} \right] \right |^2 \,,\nonumber
\end{eqnarray} \end{widetext} where we have set \begin{eqnarray}
\varphi_j = k ({\bf r_2}-{\bf r_1})\cdot {\bf \hat x_j} = kd\cos\vartheta_j
\end{eqnarray} and defined $G^{(2)}(\varphi_1,\varphi_2;t')=\mathcal G^{(2)}({\bf x_1},t;{\bf x_2},t+t')$.
The detailed derivation of Eq.~(\ref{eqn:G2allg}) is reported in
Appendix~\ref{app:G2}. For $\kappa=0$, i.e.\ in absence of coupling, $G^{(2)}$ exhibits
an interference pattern as a function of the distance $|{\bf x_2}-{\bf x_1}|$ between the
detectors. Such interference emerges from two indistinguishable paths of two-photon
emission. It has first been predicted in~\cite{Mandel} for the case of two independent
quantum sources, and generalized in~\cite{Skornia} for the light scattered by two trapped
atoms illuminated by a laser. In particular, the result of~\cite{Skornia} for weak laser
intensity is obtained from Eq.~(\ref{eqn:G2allg}) by taking the limit $\kappa\to 0$ and
setting $\Delta=0$~\cite{Footnote}.

We now consider this spatial interference pattern at $t'=0$, for
$\varphi_L=0$ and $\Delta=0$, but keeping $\kappa\neq 0$. For
these parameters it takes the form
\begin{equation}\label{eqn:G2v0} G^{(2)}(\varphi_1,\varphi_2;0)=\frac{64 \Omega^4}{\gamma^4}~ \frac{\cos^2 (\varphi_1-\varphi_2
)/2}{1+\kappa^2+2\kappa \cos \omega_0 \tau }.
\end{equation} In particular, Eq.~(\ref{eqn:G2v0}) vanishes for
$|\varphi_1-\varphi_2|=(2n+1)\pi$, showing strong antibunching at
these points. Similarly, bunching is encountered whenever the
condition $|\varphi_1-\varphi_2|=2n\pi$ is fulfilled. We note,
moreover, that since this signal depends only on the difference
$\varphi_1-\varphi_2$, there exists a finite probability of
measuring two photons simultaneously at the positions of the
screen which are the dark fringes of the first-order interference
pattern, $\varphi_j=(2n+1)\pi$. This behaviour has been discussed
in~\cite{Skornia-2}: it is connected to the fact that saturation
effects diminish the contrast of the first-order correlation
function, leading to a non-vanishing probability of measuring a
photon at these detector positions. The probability to measure the
first photon in the dark fringe is essentially proportional to the
occupation of the collective state $|e_1,e_2\rangle$, and the
first detection projects the atoms into the antisymmetric Dicke
state $\ket{\psi(0)}=\frac{1}{\sqrt{2}}(\ket{e,g}-\ket{g,e})$,
which is an entangled state of the two distant atoms.

The denominator of Eq.~(\ref{eqn:G2v0}) shows how the spatial
interference pattern is modified due to the atom-atom interaction
by multiple photon scattering, and how this modification depends
on the phase $\omega_L\tau$.

\begin{figure}[htp]
 \centering
 \includegraphics[width=.4\textwidth] {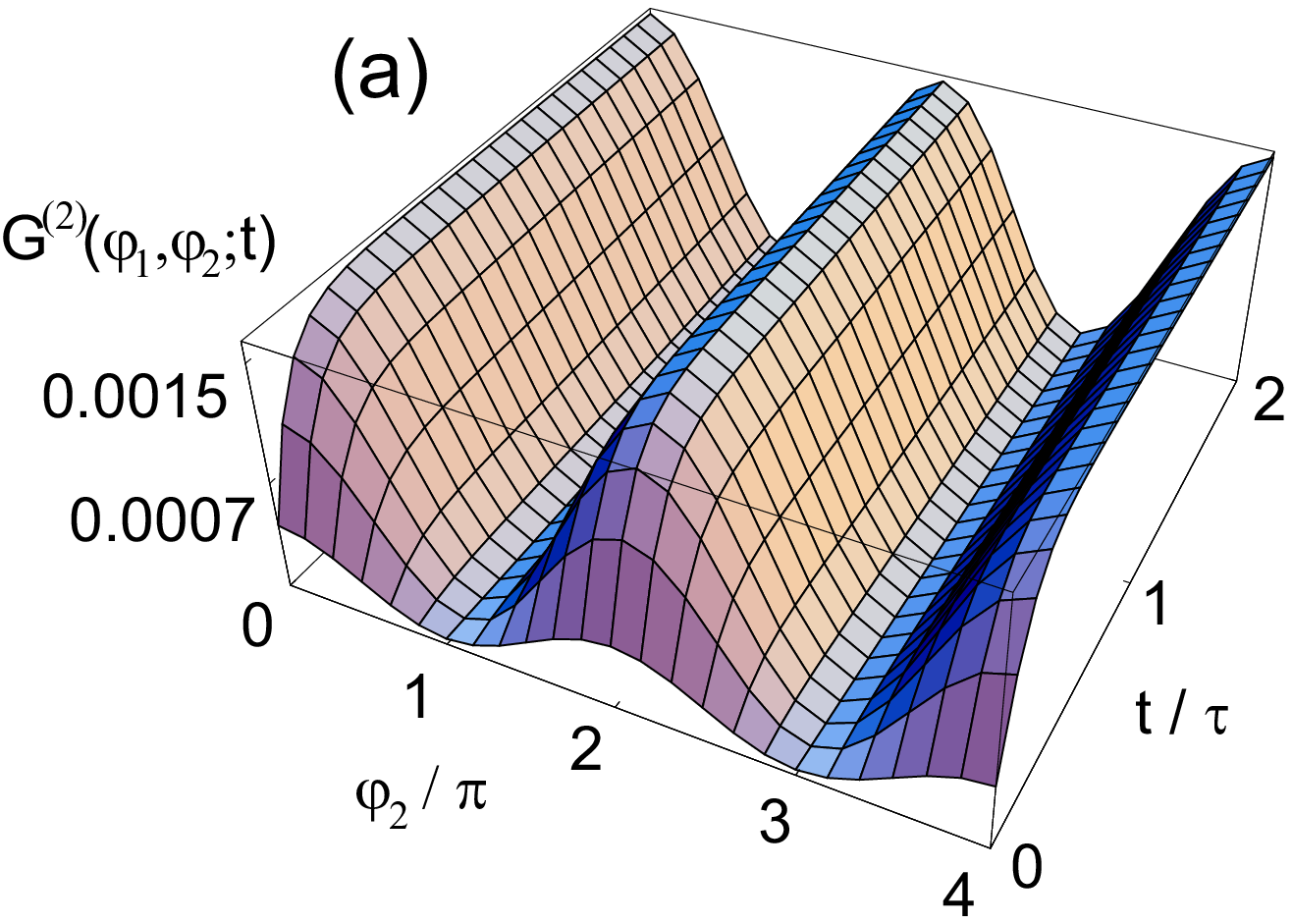}
 \includegraphics[width=.4\textwidth]{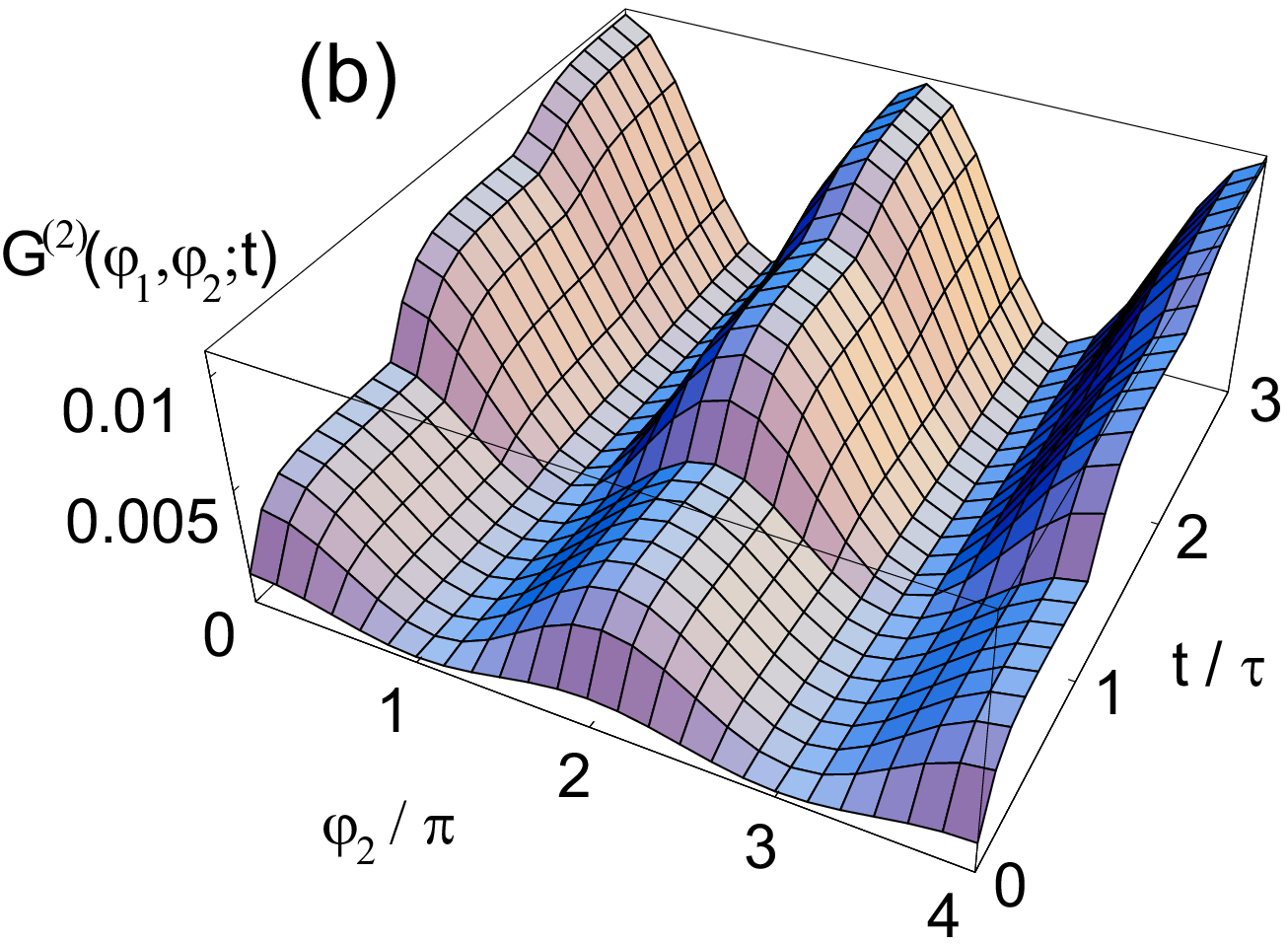}
\caption{(Color online) $G^{(2)}(\varphi_1,\varphi_2;t)$ as a function of time $t$ (in
units of $\tau$) and $\varphi_2$ for $\varphi_1=2n \pi$. The subplots refer to the cases
(a) $\kappa=0$ (no interaction) and (b) $\kappa=0.4$. The other parameters are
$\Omega=0.05\gamma$, $\Delta=0$, $\varphi_L=0$, $\gamma \tau =20$, and $\omega_0
\tau=(2n+1)\pi$.} \label{fig:G2twoD:2pi}
\end{figure}

\begin{figure}[htp]
 \centering
 \includegraphics[width=.4\textwidth]{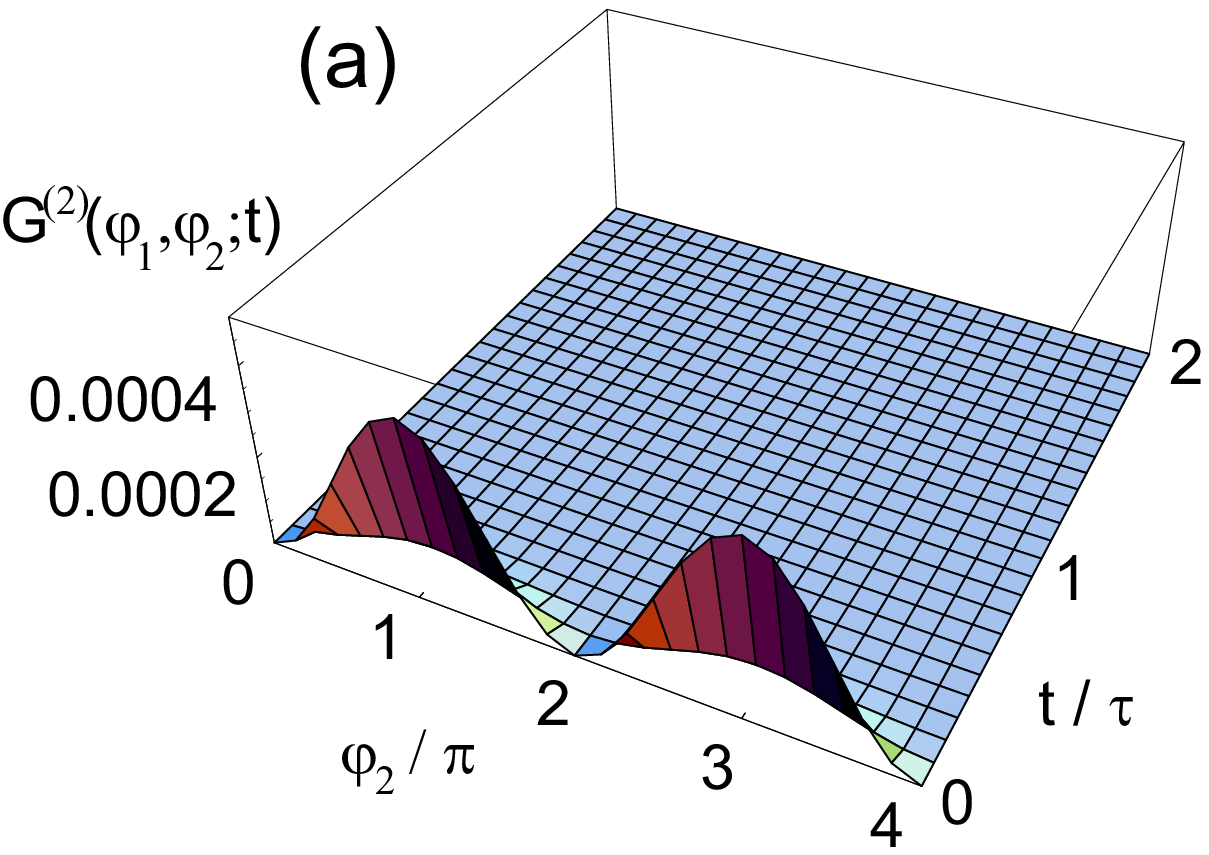}
 \includegraphics[width=.4\textwidth]{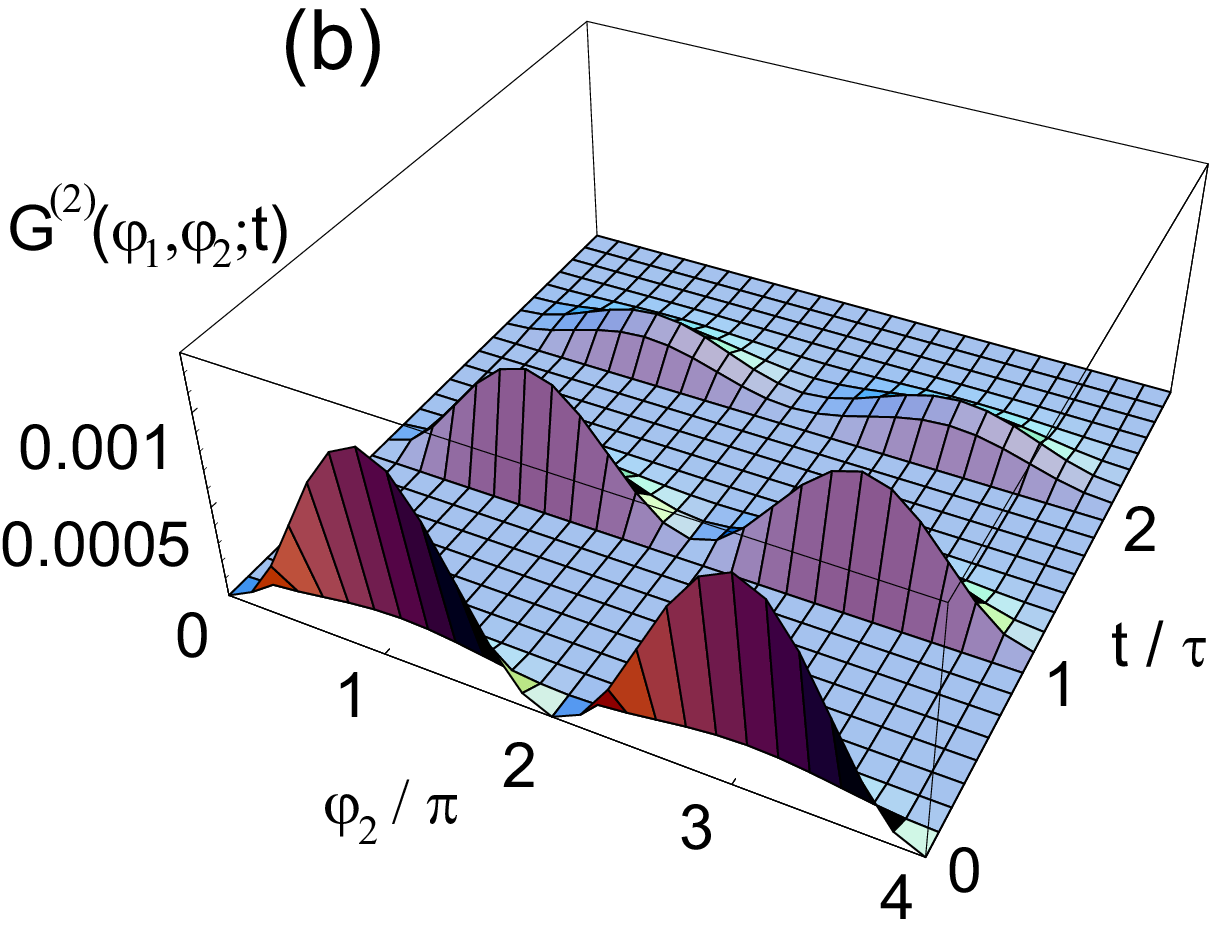}
\caption{(Color online) Same as Fig.~\ref{fig:G2twoD:2pi} but with $\varphi_1=(2n+1)\pi$.
The curves in (b) at $t>\tau$ are magnified by a factor 30.} \label{fig:G2twoD:2pi1}
\end{figure}

Figures~\ref{fig:G2twoD:2pi} and~\ref{fig:G2twoD:2pi1} display the
intensity-intensity correlation function versus $t'$ and
$\varphi_2$, for the situation $\gamma\tau\gg 1$. The two figures
correspond to the bright ($\varphi_1=2n\pi$) and dark
($\varphi_1=(2n+1)\pi$) fringes of the first-order correlation
function, and both show the cases $\kappa=0$ and $\kappa=0.4$, for
comparison. One clearly observes the effect due to multiply
scattered photons, giving rise to abrupt changes in the slope of
the correlation function at multiples of $\tau$. Hence, the
interference due to multiple scattering enhances or suppresses the
probability of measuring the second photon at a certain time
interval. Related effects have been observed in a single-atom
interference experiment in Ref.~\cite{Dubin07}. We now analyze
this latter property setting $\varphi_1=(2n+1)\pi$, i.e.\ when the
first detector is set at a dark fringe of the first-order
correlation function: in Fig.~\ref{fig:G2twoD:2pi1}(a) one sees
that for $\kappa=0$ the second order correlation function is
different from zero at $t'=0$ and for $\varphi_2\neq 2n\pi$, and
it vanishes after a transient time of the order $1/2\gamma$,
corresponding to the lifetime of the collective state
$|e,e\rangle$~\cite{Skornia-2}. For $\kappa=0.4$,
Fig.~\ref{fig:G2twoD:2pi1}(b), one observes "revivals" when the
time between the two detections is a multiple of $\tau$, and whose
amplitude is strongly damped as a function of time. Inspecting
Eq.~(\ref{eqn:G2allg}) for these specific parameters, we find that
at short times the correlation function behaves as
\begin{eqnarray}
&&G^{(2)}(\varphi_1,\varphi_2;t')\Bigl|_{\varphi_1=(2n+1)\pi,t'\ll \tau}\approx\\
&& \frac{32 \Omega^4 (1-\cos \varphi_2)}{\gamma^4 (1+\kappa^2+2\kappa \cos \omega_0 \tau
)} \left | \sum_k (-1)^k I_k  \right |^2 \,,\nonumber \end{eqnarray} and it is essentially
proportional to the probability of measuring the atoms in the state $\ket{\psi(t')}$,
obtained by freely evolving the initial state
$\ket{\psi(0)}=\frac{1}{\sqrt{2}}(\ket{e,g}-\ket{g,e})$, according to
Eq.~(\ref{eqn:excitedOhneLaser}).

For longer times, $t'>\tau$, the second-order correlation function
scales with $\kappa^2$ and takes the form
\begin{eqnarray}\label{eqn:G2oneAtom}
&&G^{(2)}(\varphi_1,\varphi_2;t') \Bigl|_{\varphi_1=(2n+1)\pi,t'>\tau}\\
&&\approx \frac{64 \Omega^4 (1-\cos^2 \varphi_2)} {\gamma^4 (1+\kappa^2+2\kappa \cos
\omega_0 \tau )} \Bigl| K\sum_k H_k(\omega_L,t') \Bigr|^2 \,,\nonumber \end{eqnarray} which
is essentially proportional to the stationary excited-state occupation of the atoms given
in Eq.~(\ref{eqn:MLexcited}).

\begin{figure}[htp]
 \centering
 \includegraphics[width=.45\textwidth]{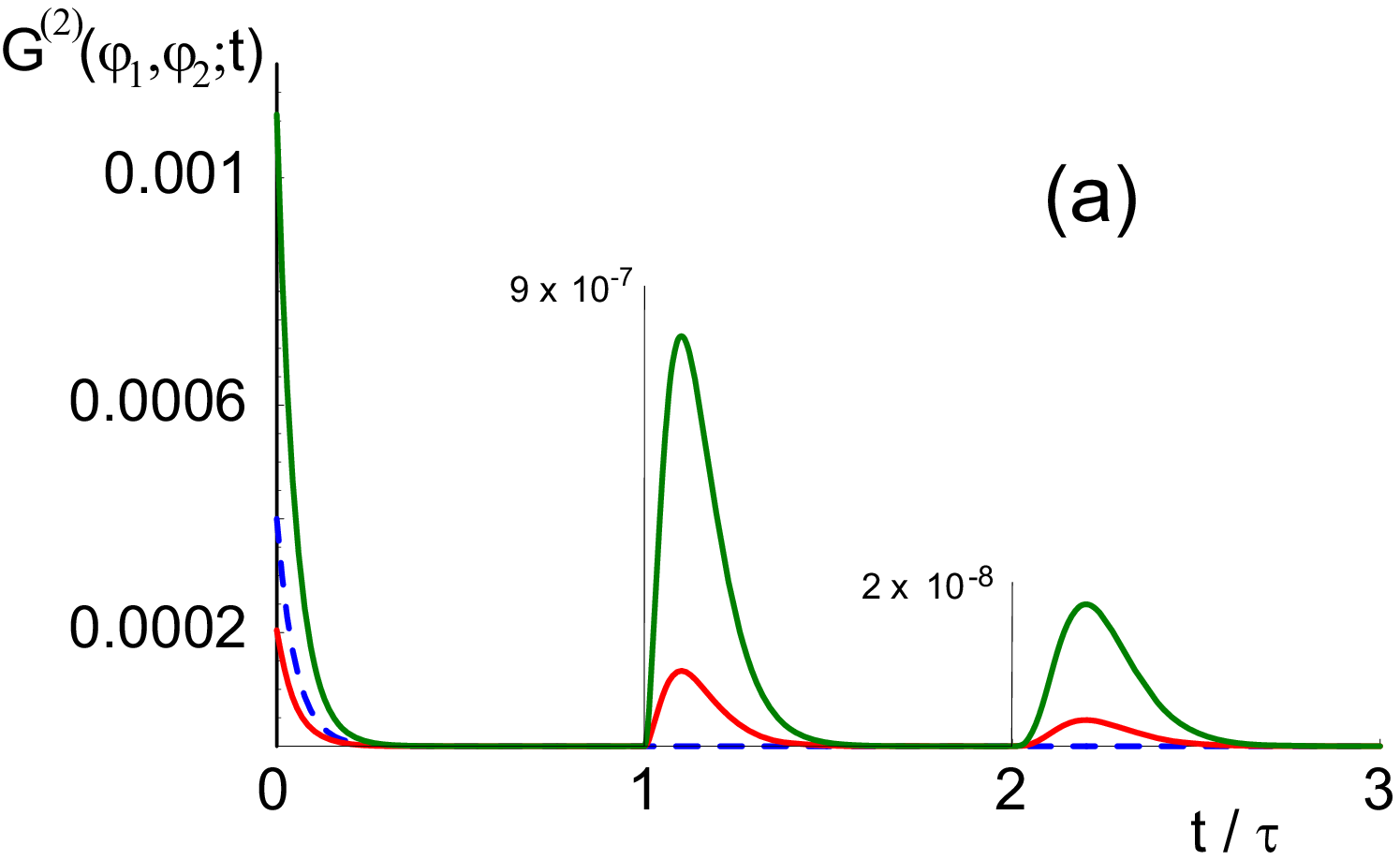}
 \includegraphics[width=.45\textwidth]{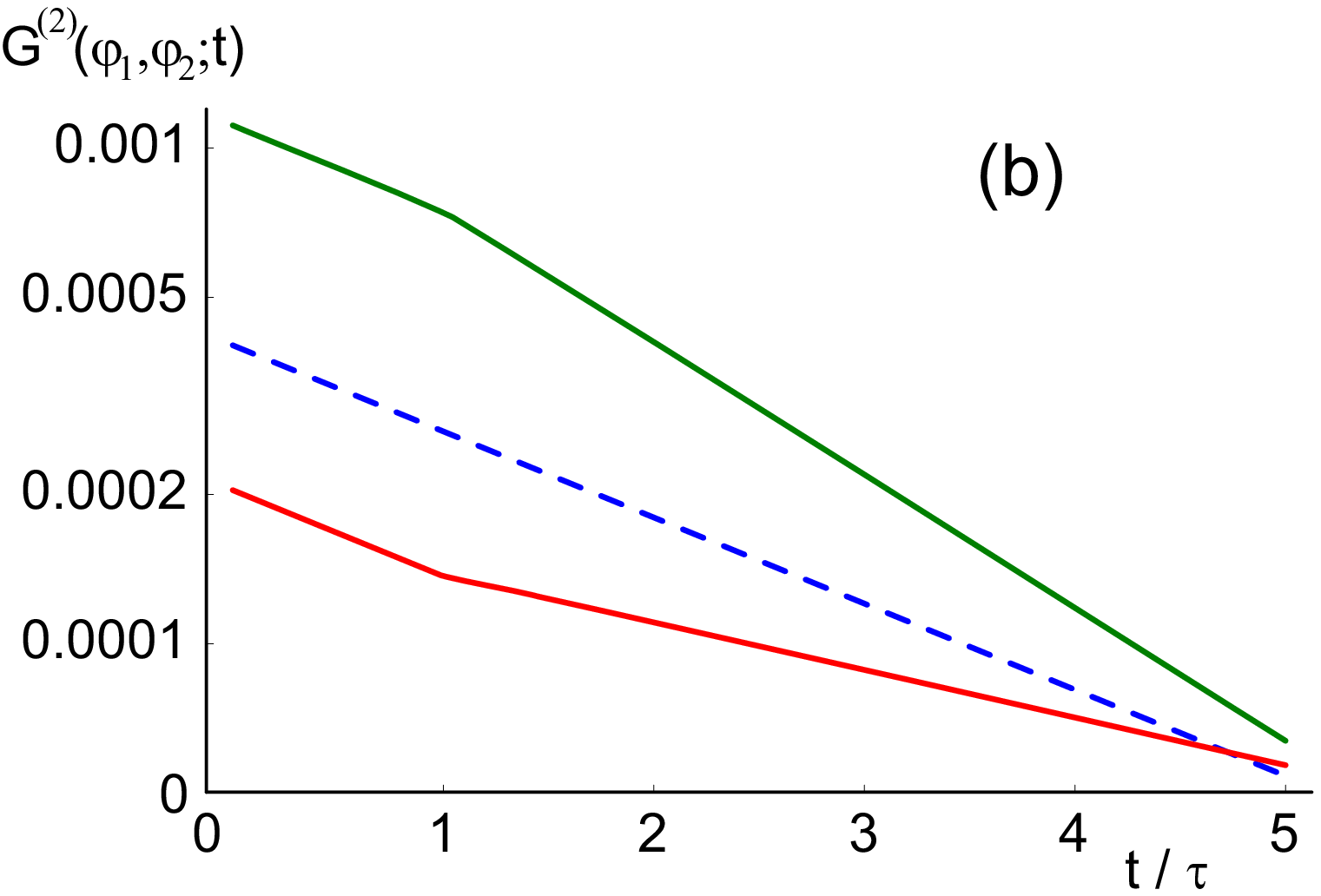}
\caption{(Color online) $G^{(2)}(\varphi_1,\varphi_2;t)$ as a function of time for
$\varphi_1=\varphi_2=(2n+1) \pi$ and $\kappa=0.4$ when (a) $\gamma\tau=20$ (note the
change of vertical scale from each maximum to the next one) and (b) $\gamma\tau=0.4$. The
red-solid (bottom) and green-solid (top) curves are evaluated at $\omega_0 \tau = 2n\pi$ and
$(2n+1)\pi$, respectively. The blue dashed line represents the behaviour at $\kappa=0$
and is plotted for reference. The other parameters are $\Omega=0.05\gamma$, $\Delta=0$,
$\varphi_L=0$. } \label{fig:G2twoD:single}
\end{figure}

While the limit $\gamma\tau\gg 1$ is characterized by "revivals"
of the correlation function versus the time $t'$ between two
photon detections, in the limit $\gamma\tau\le 1$ one observes a
smooth decay of the correlation function with $t'$, whereby its
decay rate is modified depending on whether the multiply scattered
waves interfere constructively or destructively at the atom. A
comparison between the two regimes is shown in
Fig.~\ref{fig:G2twoD:single}.

To conclude this section, one of the main features associated with
photon-mediated atom-atom interaction is that the
intensity-intensity correlation exhibits an enhanced or suppressed
probability to measure a second photon as a function of the time
after the first detection. This behaviour is due to interference
between the various paths of multiple scattering, and can be
interpreted as a combined photonic-atomic excitation which is
stored inside the system. In view of the interpretation
by~\cite{Skornia-3,Metz}, one can say that for a transient time
the system develops and stores entanglement and correlations,
determined by the strength of the interaction $\kappa$, until the
atoms finally dissipate the excitation into free space. In the
future, it would be interesting to consider these dynamics in a
quantum jump picture~\cite{Skornia-3}. This could open the
possibility of implementing schemes for entangling atoms in this
kind of set-up, as proposed in~\cite{Metz}.

\section{Conclusions} \label{Sec:Conclusion}

We have studied the photonic properties of two atoms which are
coupled by radiation via an optical element such as, e.g., a lens
or an optical fiber focussing a relevant fraction $\kappa$ of
electromagnetic field modes from one atom to the other. Signatures
of multiple scattering of photons between the atoms are observed
in the first-order and second-order coherence of the scattered
light. These features show that the presence of the second atom
substantially modifies the radiative properties of the first one,
even when the atoms are separated by a distance $d$ much larger
than the light wavelength $\lambda$.

The efficiency of the interaction, which for two atoms in free
space scales with $\lambda/d$, is determined by $\kappa$ when an
optical system mediates the coupling. The atom-atom distance $d$
plays a new role, separating two regimes where the delay of the
interaction $\tau=d/c$ is smaller or larger than the atomic decay
time $1/\gamma$. In these two regimes $\gamma\tau<1$ and
$\gamma\tau>1$, the coupled two-atom system shows characteristics
of a single- or multi-mode resonator, respectively, with mirrors
of low reflectivity $\kappa$ and bandwidth $\gamma$.

In this article we considered the limit in which the atoms are
weakly driven by the laser, and we neglected the effect of atomic
motion. It should be remarked that localization of the atoms
within a wavelength of the scattered light is a relevant
requirement for observing the interference effects arising from
multiple scattering. In fact, atomic motion gives rise to a
dephasing in the signal, which can be interpreted as which-way
information imprinted by the photon recoil on the scattering
atom~\cite{Wickles,Eschner2003,Englert}. Nevertheless, when the
recoil of the atom in each photon scattering event is taken into
account, correlations between the atomic motion and the light are
established~\cite{Eberly}. In particular, mechanical effects
between the distant atoms arise, which are mediated and retarded
by the optical coupling~\cite{Bushev2004, iacopini93}. It is
interesting to consider whether such effects may lead to novel
collective behaviour of atomic center-of-mass and photonic
variables, in analogy to collective dynamics predicted for cold
atoms inside resonators~\cite{Ritsch}.

\acknowledgements

The authors acknowledge discussions with and helpful comments of
Endre Kajari, Georgina Olivares-Renteria, and Wolfgang Schleich.
This work was supported by the European Commission (EMALI,
MRTN-CT-2006-035369; Integrated Project SCALA, Contract No.\
015714) and by the Spanish Ministerio de Educaci\'on y Ciencia
(Consolider-Ingenio 2010 QOIT, CSD2006-00019; QLIQS,
FIS2005-08257; QNLP, FIS2007-66944; Ramon-y-Cajal; Acci{\'o}n
Integrada HA2005-0001).

\begin{appendix}

\section{} \label{App:A}

We formally integrate Eqs.~(\ref{eqn:koefffield:rho}) and~(\ref{eqn:koefffield}) using the initial condition
$b_g^{(\ell)}(0)=0$. Inserting the result into Eq.~(\ref{1}) yields
\begin{eqnarray} \label{eqnb1pkt}
&&\dot{b}_e^{(1)}(t) =-\sum_{\ell=\mu,\rho} g_{\ell}^2
                \int_0^t {\rm d}t' e^{{\rm i} (\omega_0 -\omega_{\ell})(t-t')} b_e^{(1)}(t')\\
&&-\sum_{\rho} g_{\rho}^2
                \int_0^t {\rm d}t' e^{{\rm i} (\omega_0 -\omega_{\rho})(t-t')} {\rm e}^{{\rm i}{\bf k_{\rho}}
                \cdot ({\bf r_1}-{\bf r_2})} b_e^{(2)}(t')\,,\nonumber
\end{eqnarray}  where the equation
for $b_e^{(2)}(t)$ is found by swapping the indexes $1$ and $2$ in
Eq.~(\ref{eqnb1pkt}). We see that the differential equation for
the probability amplitude $b_e^{(1)}(t)$ depends linearly on the
probability amplitude $b_e^{(2)}(t')$ for the excited state of
atom $2$. Such dependence is due to the common modes $\rho$ which
mediate the interaction between the two atoms. In absence of the
second atom, the equation reduces to the well known equations
describing the
radiative decay of a two-level atom in free space~\cite{Cohen-Tannoudji}.\\

The first term on the right-hand side (RHS) of Eq.~(\ref{eqnb1pkt}) can be rewritten in a compact way by converting the sum over the modes into an integral. Using the Wigner-Weisskopf approximation one obtains~\cite{Milonni}
\begin{eqnarray}\label{eqn:1}
&&\dot{b}_e^{(1)}(t) = -\frac{\gamma}{2} b_e^{(1)}(t) \\
&&-\int_0^t {\rm d}t' b_e^{(2)}(t')  \sum_{\rho} g_{\rho}^2 {\rm
e}^{{\rm i}{\bf k_{\rho}}\cdot ({\bf r_1}-{\bf r_2})}{\rm e}^{{\rm
i} (\omega_0-\omega_{\rho}) (t-t')}\,,\nonumber \end{eqnarray}
 where $\gamma =D^2 \omega_0^3/(3 \pi
\varepsilon_0 \hbar c^3)$ is the free-space decay
rate~\cite{Cohen-Tannoudji}. The second term on the RHS of
Eq.~(\ref{eqn:1}) corresponds to the sum of the modes mapped from
one atom into the other by the optical setup. Let us consider a
lens between the atoms collecting a solid angle $\delta\Omega_0$
of modes with aperture $\theta_0$. Converting the sum over the
modes into an integral, the sum over the modes $\rho$ in
Eq.~(\ref{eqn:1}) can be rewritten as \begin{eqnarray}
&&\sum_{\rho} g_{\rho}^2 {\rm e}^{{\rm i}{\bf k_{\rho}}\cdot ({\bf r_1}-{\bf r_2})}{\rm e}^{{\rm i} (\omega_0-\omega_{\rho}) (t-t')}\to\nonumber\\
&&\frac{1}{2(2 \pi)^3\varepsilon_0 \hbar c^3} \int_0^{\infty} {\rm d}\omega \omega^3 {\rm e}^{{\rm i} (\omega_0-\omega) (t-t')} \nonumber \\
&&\times \int_{\delta\Omega_0}{\rm d}\Omega {\rm e}^{{\rm i}{\bf
k}\cdot ({\bf r_1}-{\bf r_2})} \left (D^2-\frac{|{\bf D}\cdot {\bf
k}|^2}{k^2} \right )\,.\label{eqn:mirrorlens} \end{eqnarray} Since
optics compensate for the phase difference between the various
modes, we take \begin{equation} {\rm e}^{{\rm i}{\bf k}\cdot ({\bf
r_1}-{\bf r_2})}\rightarrow e^{{\rm i} \omega \tau}\,, \end{equation}
where $\tau$ is defined in Eq.~(\ref{tau}), and is the time a
photon emitted inside the solid angle $\delta\Omega_0$ needs to
cover the distance between one atom and the other via optical
setup. Using Eq.~(\ref{eqn:mirrorlens}) into Eq.~(\ref{eqn:1}), we
can now make the Wigner-Weisskopf approximation and obtain
Eq.~(\ref{eqns:b}).

\section{} \label{Sterms}

We consider the terms of Eq.~(\ref{eqn:MLfeldkoeffrech}). Let us introduce the simple relation
\begin{eqnarray}\label{eqn:MHintegral}
 & & \mathcal I=\int_0^t {\rm d}t' {\rm e}^{{\rm i} (\omega_L -\omega_{\mu} )t'} H_k(t',\omega_{\mu} ) \nonumber \\
 & & =\frac{( -\kappa \gamm )^k}{k!}{\rm e}^{{\rm i} \omega_L k\tau } \Theta (t-k\tau) \int_0^{t-k\tau} {\rm d}x {\rm e}^{{\rm i} (\omega_L -\omega_{\mu} )x} x^k G_k(\alpha x)\,,\nonumber\\
\end{eqnarray}
with $\alpha =\gamm +{\rm i}\delta_{\mu}$ and where we used Eq.~(\ref{eqn:MHdef}). Using the definition of the confluent hypergeometric function~\cite{Abramowitz},
\begin{equation}
_1F_1(a,b,z)=1+\frac{a}{b}z+\frac{a (a+1)}{b (b+1)}\frac{z^2}{2!}+\ldots\,,
\end{equation}
we rewrite Eq.~(\ref{eqn:MLDefG}) as
\begin{equation}
\label{Eq:G:ks}
G_k(s) = \left\{ \begin{array}{ll}
         1-e^{-s} & \mbox{if $k=0$};\\
         -\sum_{n=0}^{\infty} \frac{n}{n+k} \frac{(-s)^n}{n!} & \mbox{if $k \neq 0$}.\end{array} \right.
\end{equation} or equivalently
\begin{eqnarray}\label{eqn:Gk:s} G_k(s) &=&\frac{k!}{s^k}-(-1)^k s
\frac{\partial^k}{\partial s^k} \frac{e^{-s}}{s}\,, \end{eqnarray}
whereby \begin{equation} \label{B:Rel:1} (-1)^k s
\frac{\partial^k}{\partial s^k} \frac{e^{-s}}{s}={\rm
e}^{-s}\sum_{n=0}^k\frac{k!}{(k-n)!}s^{-n}. \end{equation} We use
Eq.~(\ref{eqn:Gk:s}) in Eq.~(\ref{eqn:MHintegral}), and obtain
\begin{eqnarray} \label{Eq:Integral} \mathcal I=\frac{( -\kappa
\gamm{\rm e}^{{\rm i} \omega_L \tau } )^k}{\alpha^k} \Theta
(t-k\tau) \left(\int_0^{t-k\tau} {\rm d}x {\rm e}^{{\rm i}
(\omega_L -\omega_{\mu} )x} -S_k\right)\,, \end{eqnarray} with
\begin{equation} \label{Eq:S:k} S_k(t) =
\frac{(-1)^k}{k!}\int_0^{t-k\tau} {\rm d}x {\rm e}^{-{\rm i}
\Delta_{\mu}x} x^{k+1} \partiell{x}{k} \frac{e^{-\alpha x}}{x}\,.
\end{equation} As we are interested in evaluating the scattering
processes for long times, $t\to \infty$, we neglect the term
$k\tau$ in the upper bound of the integral and take the Heaviside
function to be one. The first term inside the parentheses on the
RHS of~Eq.(\ref{Eq:Integral}) gives \begin{eqnarray} \int_0^t
{\rm d}x \, {\rm e}^{-{\rm i} \Delta_{\mu} x} &=& {\rm e}^{-{\rm
i}\Delta_{\mu}t/2}   2\pi \delta^{(t)}(\Delta_{\mu})\,. \end{eqnarray}
In
order to evaluate the term~(\ref{Eq:S:k}) we use the relations
Eq.~(\ref{B:Rel:1}) and \begin{eqnarray*} \int_0^t {\rm d}x \, x^j
{\rm e}^{-\alpha x} & =& \frac{j!}{\alpha^{j+1}}-\frac{j! {\rm
e}^{-\alpha t}}{\alpha^{j+1}}\sum_{l=0}^j \frac{(\alpha t)^l}{l!}\,,
\end{eqnarray*} in Eq.~(\ref{Eq:S:k}), which then reads
\begin{eqnarray*}
S_k(t) &=& \sum_{j=0}^k \frac{\alpha^j}{(\alpha+{\rm i} \Delta_{\mu})^{j+1}}  \\
  & & \times \left(1-{\rm e}^{-(\alpha+{\rm i} \Delta_{\mu} )t} \sum_{l=0}^j \frac{(\alpha+{\rm i} \Delta_{\mu} )^l t^l}{l!}\right)\,. \end{eqnarray*}
In particular,
\begin{equation}
\sum_{k=0}^{\infty} \begin{array}{l}
                        K'^{2k} S_{2k} \\
                        K'^{2k+1} S_{2k+1}
                     \end{array} = \frac{1}{2} \left [ \sum_{k=0}^{\infty} K'^k S_k \pm (-1)^k  K'^k S_k \right ]\,,
\end{equation}
where
\begin{equation}
\sum_{k=0}^{\infty} (\pm 1)^k K'^k S_k \approx  \frac{1}{(1\mp K')(\alpha+{\rm i} \Delta_{\mu} \mp \alpha K' )}\,,
\end{equation}
with $K'=-\kappa\gamm {\rm e}^{{\rm i} \omega_L \tau}/(\gamm +{\rm i}\delta_{\mu})$, and where we used the Cauchy-Product for absolute convergent series, thereby neglecting the vanishing exponentials as we consider the long time limit.\\ 
In the long-time limit, using that $\lim_{t \to +\infty}
\delta^{(t)}(x)=\delta (x)$ we finally arrive to the relation in
Eq.~(\ref{eqn:MLfeldcoessmitS}).

\section{} \label{app:G2}

Starting from Eq.~(\ref{Eq:G2}) we calculate the second order correlation function of two
atoms, which are weakly driven by the laser and both scatter towards the detector. In the
reference frame rotating at the laser frequency, assuming that the initial state is the
atomic ground state, we rewrite Eq.~(\ref{Eq:G2}) as \begin{eqnarray}\label{eqn:G2}
&&{\mathcal G}^{(2)}({\bf
x_1},t;{\bf x_2},t+t') = \\
&&\|(\sg_1 +\sg_2 e^{{\rm i} \varphi_2}) U(t' ) (\sg_1 +\sg_2 e^{{\rm i} \varphi_1} )
U(t) \ket{g,g,0} \|^2\,, \nonumber\end{eqnarray} where the correlation function is evaluated
at lowest order in perturbation theory in the atom-photon interactions. The operator
$U(t)$ is the total evolution operator, $$U(t)=\exp(-{\rm i}H' t/\hbar)\,,$$ with $H'$ given
in Eq.~(\ref{H:tot}), which is to be expanded in power series of the interactions $V_{\rm
emf}$ and $V_L$. At lowest, non-vanishing order, Eq.~(\ref{eqn:G2}) can be rewritten as
\begin{eqnarray} \label{eqn:G2app} & & {\mathcal G}^{(2)}({\bf
x_1},t;{\bf x_2},t+t') \nonumber \\
 &&= \left | \left ( A_1(t)+A_2(t) e^{{\rm i} \varphi_1} \right ) \left ( A_1(t' )+A_2( t' ) e^{{\rm i} \varphi_2} \right )+ B(t) \right . \nonumber \\
 && \times \left .  \left [ C_{11}(t') + {\rm e}^{{\rm i} \varphi_1}C_{21}(t') + {\rm e}^{{\rm i} \varphi_2} \left(
 C_{12}(t') + C_{22}(t')  {\rm e}^{{\rm i} \varphi_1} \right )   \right ] \right |^2 \,,\nonumber \\
\end{eqnarray} where the coefficients $A_j$ and $B$ are the
transition amplitudes \begin{eqnarray} &&A_1(t)=\langle
e,g,0|U(t)|g,g,0\rangle=c_e^{(1)}(t)\,,\\
&&A_2(t)=\langle
g,e,0|U(t)|g,g,0\rangle=c_e^{(2)}(t)\,,\\
&&B(t)=\langle e,e,0|U(t)|g,g,0\rangle \,,\end{eqnarray} where
$c_{e}^{(j)} (t)$ is given in Eq.~(\ref{eqn:MLexcited}), while the
probability amplitudes $C_{ji}$ are defined as \begin{eqnarray*}
C_{11}&=&\langle e,g,0|U(t)|e,g,0\rangle \nonumber\\&=& e^{-{\rm
i} \Delta t} b_e^{(1)} (t) \hspace{0.5cm} \mbox{with}
\hspace{0.5cm} b_e^{(1)}(0)=1\,,\\
C_{12}&=&\langle g,e,0|U(t)|e,g,0\rangle \nonumber\\&=& e^{-{\rm
i} \Delta t} b_e^{(2)} (t) \hspace{0.5cm} \mbox{with}
\hspace{0.5cm} b_e^{(1)}(0)=1\,,\nonumber\\ C_{21}&=&\langle
e,g,0|U(t)|g,e,0\rangle \nonumber\\&=& e^{-{\rm i} \Delta t}
b_e^{(1)} (t) \hspace{0.5cm} \mbox{with}
\hspace{0.5cm} b_e^{(2)}(0)=1\,,\\
C_{22}&=&\langle g,e,0|U(t)|g,e,0\rangle \nonumber\\&=& e^{-{\rm
i} \Delta t} b_e^{(2)} (t) \hspace{0.5cm} \mbox{with}
\hspace{0.5cm} b_e^{(2)}(0)=1\,, \end{eqnarray*} with $b_e^{(j)} (t)$
given in Eq.~(\ref{eqn:excitedOhneLaser}). We evaluate the
coefficient $B(t)$ in second-order perturbation theory, hence
obtaining \begin{eqnarray} B(t) &=& -{\rm i}\Omega  \int_0^t dt'
{\rm  e}^{-(2{\rm i} \Delta +\gamma) (t-t')} \nonumber \\ & &
\times \left({\rm e}^{{\rm i}{\bf k_L}\cdot {\bf
r_2}}c_{e}^{(1)}(t')+{\rm e}^{{\rm i}{\bf k_L}\cdot {\bf r_1}}
c_{e}^{(2)}(t')\right) \,.\end{eqnarray} Inserting the explicit value
of the coefficients, Eq.~(\ref{eqn:MLexcited}), after partially
integrating $B(t)$ reads \begin{eqnarray} B(t) &=&
-\frac{\Omega^2}{\alpha^2}\Bigl[2{\rm e}^{{\rm i}{\bf k_L}\cdot
({\bf r_1}+{\bf r_2})}\sum_{k}F_{2k}\\
& &+\left({\rm e}^{2{\rm i}{\bf k_L}\cdot {\bf r_1}}+{\rm
e}^{2{\rm i}{\bf k_L}\cdot {\bf
r_2}}\right)\sum_{k}F_{2k+1}\Bigr]\,,\nonumber \end{eqnarray} where
\begin{eqnarray} F_k&=& \left(\gamm\frac{-\kappa{\rm e}^{{\rm
i}\omega_L\tau}}{\alpha}\right)^{k}\frac{1}{k!} \Theta(t-k\tau )
\Bigl \{ \Gamma^{} (k+1,\alpha (t-k\tau ))
\nonumber \\
& &+ (-1)^k {\rm e}^{-2 \alpha(t- k \tau) } \Gamma(k+1,-\alpha
(t-k\tau )) \Bigr \}\,, \nonumber\end{eqnarray} with $\alpha = \gamm
+{\rm i}\Delta $ and $\Gamma(k,\alpha)$ the generalized gamma
function~\cite{Abramowitz}. In the long-time limit the second term
in the curly brackets is negligible, and we find \begin{eqnarray}
B(t) &=& \frac{- \Omega^2 }{2(\gamm +{\rm
i}\Delta)^2}\frac{1}{1-K^2}\\
& &\times \left [ 2{\rm e}^{{\rm i}{\bf k_L}\cdot ({\bf r_1}+{\bf
r_2})}-K\left({\rm e}^{2{\rm i}{\bf k_L}\cdot {\bf r_1}}+{\rm
e}^{2{\rm i}{\bf k_L}\cdot {\bf r_2}}\right)\right ]\,,\nonumber
\end{eqnarray}  while the form $A_j(t)$ valid in the long-time
limit is given by Eq.~(\ref{c:e:1:long}). Inserting the explicit
value of the coefficients in Eq.~(\ref{eqn:G2app}) we finally
obtain Eq.~(\ref{eqn:G2allg}). \end{appendix}


\begin{thebibliography}{99}

\bibitem{ZollerRoadmap}
P.\ Zoller {\it et al.}, {\it Quantum information processing and
communication}, Eur.\ Phys.\ J.\ D {\bf 36}, 203 (2005).

\bibitem{Monroe04}
B.\ B.\ Blinov, D.\ L.\ Moehring, L.\ - M.\ Duan, C.\ Monroe,
Nature {\bf 428}, 153-157 (2004).

\bibitem{Weinfurter06}
J.\ Volz, M.\ Weber, D.\ Schlenk, W.\ Rosenfeld, J.\ Vrana, K.\
Saucke, C.\ Kurtsiefer, H.\ Weinfurter, Phys.\ Rev.\ Lett.\ {\bf
96}, 030404 (2006).

\bibitem{Grangier06}
A.\ Ourjoumtsev, R.\ Tualle-Brouri, J.\ Laurat, P.\ Grangier,
Science {\bf 312}, 83 (2006).

\bibitem{Maunz}
D. L. Moehring, P. Maunz, S. Olmschenk, K. C. Younge, D. N.
Matsukevich, L.-M. Duan, and C. Monroe, Nature {\bf 449}, 68
(2007).

\bibitem{Guthoerlein2001}
G.R. Guth\"ohrlein, M. Keller, K. Hayasaka, W. Lange, and H.
Walther, Nature {\bf 414}, 49-51 (2001).

\bibitem{Mundt}
A.B. Mundt, A. Kreuter, C. Becher, D. Leibfried, J. Eschner, F.
Schmidt-Kaler, R. Blatt, Phys. Rev. Lett. {\bf 89}, 103001 (2002);
A.\ Kreuter, C.\ Becher, G.P.T.\ Lancaster, A.B.\ Mundt, C.\
Russo, H.\ H\"affner, C.\ Roos, J.\ Eschner, F.\ Schmidt-Kaler,
R.\ Blatt,  Phys.\ Rev.\ Lett.\ {\bf 92}, 203002 (2004).

\bibitem{Rauschenbeutel}
I.\ Dotsenko, W.\ Alt, M.\ Khudaverdyan, S.\ Kuhr, D.\ Meschede,
Y.\ Miroshnychenko, D.\ Schrader, A.\ Rauschenbeutel, Phys.\ Rev.\
Lett.\ {\bf 95}, 033002 (2005).

\bibitem{Walther04}
M.\ Keller, B.\ Lange, K.\ Hayasaka, W.\ Lange, H. Walther, Nature
{\bf 431}, 1075 (2004).

\bibitem{Kimble04}
J.\ McKeever, A.\ Boca, A.D.\ Boozer, R.\ Miller, J.R.\ Buck, A.\
Kuzmich, H.J.\ Kimble, Science {\bf 303}, 1992 (2004).

\bibitem{Kuhn2}
M. Hijlkema, B. Weber, H.P. Specht, S.C. Webster, A. Kuhn, and G.
Rempe, Nature Physics {\bf 3}, 253 (2007).

\bibitem{Rempe07}
T. Wilk, S. C. Webster, A. Kuhn, and G. Rempe, Science {\bf 317},
488 (2007).

\bibitem{Kimble08}
B. Dayan, A.S. Parkins, T. Aoki, E.P. Ostby, K.J. Vahala, and H.J.
Kimble, Science {\bf 319}, 1062 (2008).

\bibitem{Eschner2001}
J. Eschner, Ch. Raab, F. Schmidt-Kaler and R.Blatt, Nature {\bf
413}, 495 (2001).

\bibitem{Wilson}
M. A. Wilson, P. Bushev, J. Eschner, F. Schmidt-Kaler, C. Becher,
R. Blatt, U. Dorner, Phys. Rev. Lett. {\bf 91}, 213602 (2003);

\bibitem{Bushev2004}
P. Bushev, A.Wilson, J. Eschner, C. Raab, F. Schmidt-Kaler, C.
Becher, R. Blatt, Phys. Rev. Lett. {\bf 92}, 223602 (2004).

\bibitem{Sondermann07}
M. Sondermann, R. Maiwald, H. Konermann, N. Lindlein, U. Peschel,
and G. Leuchs, Appl. Phys. B {\bf 89}, 489 (2007).

\bibitem{Leuchs07}
N. Lindlein, R. Maiwald, H. Konermann, M. Sondermann, U. Peschel
and G. Leuchs, Laser Physics, {\bf 17}, 927-934 (2007)

\bibitem{Kurtsiefer}
Meng Khoon Tey, Zilong Chen, Syed Abdullah Aljunid, B. Chng, F.
Huber, G. Maslennikov, and C. Kurtsiefer, preprint,
arXiv:0802.3005v2 (2008).

\bibitem{Spreeuw}
V.V. Ivanov, R.A. Cornelussen, H.B. van Linden van den Heuvell,
and R.J.C. Spreeuw, J. Opt. B {\bf 6}, 454 (2004).

\bibitem{Balykin}
Fam Le Kien, S. Dutta Gupta, V.I. Balykin, and K. Hakuta, Phys.
Rev. A {\bf 72}, 032509 (2005).

\bibitem{Arno}
G. Sagu\'e, E. Vetsch, W. Alt, D. Meschede, and A. Rauschenbeutel,
Phys. Rev. Lett. {\bf 99}, 163602 (2007).

\bibitem{Itano}
U. Eichmann, J. C. Bergquist, J. J. Bollinger, J. M. Gilligan, W.
M. Itano, D. J. Wineland, and M. G. Raizen, Phys. Rev. Lett {\bf
70}, 2359 (1993);
W. M. Itano, J. C. Bergquist, J. J. Bollinger, D. J. Wineland, U.
Eichmann, and M. G. Raizen, Phys. Rev. A {\bf 57}, 4176 (1998).

\bibitem{DeVoe}
R.G. DeVoe and R.G. Brewer, Phys. Rev. Lett. {\bf 76}, 2049
(1996).

\bibitem{Alber}
G. Alber, Phys. Rev. A {\bf 46}, R5338 (1992)

\bibitem{Dorner}
U. Dorner and P. Zoller, Phys. Rev. A {\bf 66}, 023816 (2002).

\bibitem{Cohen-Tannoudji} C. Cohen-Tannoudji, J. Dupont-Roc, G.
Grynberg {\it Atom-Photon Interactions}, Wiley eds. (2004).

\bibitem{footnote:density} In this expression we omit to write the
density of states of the electromagnetic field, which is
proportional to $\omega^2$. In fact, as $\gamma\ll\omega_0$, we
can consider this factor to be constant to very good
approximation. See also~\protect\cite{Cohen-Tannoudji}.

\bibitem{Dipole-Dipole}
R. H. Dicke, Phys. Rev. 93, 99 (1954);
M. Gross and S. Haroche, Phys. Rep. 93, 301 (1982).

\bibitem{Savels07}
T. Savels, A.P. Mosk, and A. Lagendijk, Phys. Rev. Lett.
\textbf{98}, 103601 (2007).

\bibitem{Mandel}
L. Mandel, Phys. Rev. A {\bf 28}, 929 (1983).

\bibitem{Wickles}
C. Wickles and C. M\"uller, Europhys. Lett. {\bf 74}, 240 (2006).

\bibitem{Skornia}
C. Skornia, J. von Zanthier, G. S. Agarwal, E. Werner and H.
Walther, Phys. Rev. A {\bf 64}, 063801 (2001).

\bibitem{Footnote} The result in~\protect\cite{Skornia} is found
by considering the normalized intensity-intensity correlation
function \begin{eqnarray*}
& &g^{(2)}({\bf x_1},{\bf x_2};t')=G^{(2)}({\bf x_1},{\bf x_2};t')/G^{(2)}({\bf x_1},{\bf x_2},t\to\infty)\\
& &=\frac{e^{-\gamma t} \left ( (2 e^{\gamm t} -1) \cos
\frac{\delta_1}{2} \cos \frac{\delta_2}{2}+\sin
\frac{\delta_1}{2}\sin \frac{\delta_2}{2} \right )^2}{(s+\cos
\frac{\delta_1}{2}) (s+\cos \frac{\delta_1}{2})} \end{eqnarray*}
with $\delta_j=\varphi_L +\varphi_j$. Here,
$s=2(\Omega/\gamma)^2+1$ and it tends to unity in the low
saturation limit considered here, giving rise to very large values
of $g^{(2)}({\bf x_1},{\bf x_2};t')$ whenever $\delta_j
=(2n+1)\pi$, i.e., at the dark-fringe condition of the first-order
correlation function.

\bibitem{Dubin07}
F. Dubin, D. Rotter, M. Murkherjee, C. Russo, J. Eschner, and R.
Blatt, Phys. Rev. Lett. {\bf 98}, 183003 (2007).

\bibitem{Skornia-2}
G. S. Agarwal, J. von Zanthier, C. Skornia, and H. Walther, Phys.
Rev. A \textbf{65}, 053826 (2002).

\bibitem{Skornia-3}
C. Skornia, J. von Zanthier, G. S. Agarwal, E. Werner, and H.
Walther, Phys. Rev. A \textbf{64}, 053803 (2001)

\bibitem{Metz}
J. Metz, M. Trupke, and A. Beige, Phys. Rev. Lett. {\bf 97},
040503 (2006);
J. Metz and A. Beige, Phys. Rev. A {\bf 76}, 022331 (2007).

\bibitem{Eschner2003}
J. Eschner, Eur. Phys. J. D {\bf 22}, 341-345 (2003).

\bibitem{Englert}
B. G. Englert, Phys. Rev. Lett. {\bf 77}, 2154 (1996).

\bibitem{Eberly}
K.W. Chan, C.K. Law, and J.H. Eberly, Phys. Rev. A \textbf{68},
022110 (2003).

\bibitem{iacopini93}
E. Iacopini, Phys. Rev. A {\bf 48}, 129 (1993).

\bibitem{Ritsch}
P. Domokos and H. Ritsch, Phys. Rev. Lett. \textbf{89}, 253003
(2002).

\bibitem{Milonni}
P.W. Milonni and P.L. Knight, Phys. Rev. A {\bf 10}, 1096 (1974).

\bibitem{Dung}
H. T. Dung and K. Ujihara, Phys. Rev. A {\bf 59}, 2524 (1999).

\bibitem{Heitler}    
W. Heitler, {\it The Quantum Theory of Radiation} (Oxford
University Press, Oxford, 1954).

\bibitem{Abramowitz} M. Abramowitz and I.A. Stegun, {\it Handbook
of mathematical functions}, (Dover Publications Inc., New York,
1968).

\end{thebibliography}
\end{document}